\documentclass[prd,aps,amsfonts,eqsecnum,nofootinbib,longbibliography,notitlepage]{revtex4-1}

\usepackage{graphicx}
\usepackage[dvipsnames]{xcolor}
\usepackage{rotating}
\usepackage{amsmath,amssymb,graphics,amsthm,isomath}
\usepackage{amsfonts,dsfont,mathtools}
\usepackage{bbm}
\usepackage{bm}
\usepackage{array}
\newcolumntype{P}[1]{>{\centering\arraybackslash}p{#1}}
\usepackage{multirow}

\usepackage[colorlinks=true, urlcolor=violet, linkcolor=blue, citecolor=red, hyperindex=true, linktocpage=true]{hyperref}
\usepackage[capitalise,compress]{cleveref}

\allowdisplaybreaks

\numberwithin{thm}{section}

\makeatletter
\renewcommand{\p@subsection}{}
\renewcommand{\p@subsubsection}{}
\makeatother

\usepackage{xcolor}
\usepackage{mathtools}

\newcommand\bea{\begin{eqnarray}}
\newcommand\eea{\end{eqnarray}}
\newcommand\be{\begin{equation}}
\newcommand\ee{\end{equation}}
\newcommand\bes{\begin{subequations}}
\newcommand\ees{\end{subequations}}
\newcommand\bed{\begin{displaymath}}
\newcommand\eed{\end{displaymath}}
\newcommand\beal{\begin{aligned}}
\newcommand\eeal{\end{aligned}}
\newcommand\bew{\begin{widetext}}
\newcommand\eew{\end{widetext}}
\newcommand\beit{\begin{itemize}}
\newcommand\eeit{\end{itemize}}
\def\bea{\begin{array}}
\def\eea{\end{array}}
\newcommand\been{\begin{enumerate}}
\newcommand\eeen{\end{enumerate}}

\newcommand{\intxyt}{\int \mathrm{d}x \mathrm{d}y \mathrm{d}t}
\newcommand{\intxt}{\int \mathrm{d}x \mathrm{d}t}
\newcommand{\intxyzt}{\int \mathrm{d}x \mathrm{d}y \mathrm{d}z \mathrm{d}t }

\def\i{\text{i}}

\usepackage{verbatim}

\usepackage{dsfont}

\begin{document}

\title{Anomalous hydrodynamics with triangular point group in $2+1$ dimensions}

\author{Marvin Qi}
\email{marvin.qi@colorado.edu}
\affiliation{Department of Physics and Center for Theory of Quantum Matter, University of Colorado, Boulder CO 80309, USA}

\author{Jinkang Guo}
\affiliation{Department of Physics and Center for Theory of Quantum Matter, University of Colorado, Boulder CO 80309, USA}

\author{Andrew Lucas}
\email{andrew.j.lucas@colorado.edu}
\affiliation{Department of Physics and Center for Theory of Quantum Matter, University of Colorado, Boulder CO 80309, USA}

\date{\today}

\begin{abstract}
We present a theory of hydrodynamics for a vector U(1) charge in 2+1 dimensions, whose rotational symmetry is broken to the point group of an equilateral triangle.  We show that it is possible for this U(1) to have a chiral anomaly.  The hydrodynamic consequence of this anomaly is the introduction of a ballistic contribution to the dispersion relation for the hydrodynamic modes.   We simulate classical Markov chains and find compelling numerical evidence for the anomalous hydrodynamic universality class.  Generalizations of our theory to other symmetry groups are also discussed.
\end{abstract}

\maketitle

\tableofcontents

\section{Introduction}
Recent years have seen renewed interest in understanding hydrodynamics as an effective field theory.  On the one hand, this is inspired by explicit geometric constructions of the Schwinger-Keldysh dissipative action that describes the Navier-Stokes equations, and a thorough understanding of how to incorporate subtle symmetries, such as time-reversal via Kubo-Martin-Schwinger invariance \cite{crossley_effective_2017, haehl_fluid_2016, Jensen_2018}.  On the other hand, there are a variety of exotic fluids, arising in (or at least inspired by) quantum matter. Anomalies lead to clear signatures even within classical hydrodynamics \cite{Son_2009, Glorioso_2019,luca}, while electron liquids may have reduced spatial symmetries which lead to unconventional transport coefficients \cite{caleb, Friedman_triangular, Huang_2022,viscometry,anisoHydro1,anisoHydro2, anisoHydro3,anisoHydro4}.  Most recently, kinetically constrained ``fracton hydrodynamics" have been intensely studied \cite{Gromov_2020, breakdown_fracton_hydro,morningstar,knap2020,zhang2020,IaconisVijayNandkishore,iaconis2021,doshi,knap2021,Grosvenor:2021rrt,osborne,Burchards:2022lqr,quasiconservation,sala2021dynamics,Guo:2022ixk}.

In a complementary thread of research, a series of papers over the past few years \cite{vectorglobalsym, SeibergShao2d, SeibergShao3d,Gorantla:2020xap,Rudelius:2020kta,subsysteminflow,You:2021tmm,you,Gorantla:2022eem} has posed a simple question: what happens when a quantum field theory has an unusual global symmetry?  For example, suppose that there is a U(1) symmetry on each \emph{plane} of a three-dimensional cubic lattice.  The resulting subsystem symmetry can have peculiar consequences including UV-IR mixing and other subtle lattice dependences in continuum quantum field theory.  A particularly important structure which arises in these constructions is the presence of charges and or current which transform in unusual irreducible representations of the spatial rotational symmetry (usually a discrete group).  For example, in the model of planar subsystem symmetry in three dimensions, one writes down a conserved current in a three-dimensional representation $(J_{xy}, J_{yz}, J_{zx})$ of the cubic point group, descending from the spin-2 representation of SO(3).

This paper was first inspired by a simple question: what is the landscape of hydrodynamic theories that are possible when one considers a charge density $\rho^a$ and a spatial charge current $J^\alpha$ that transform in exotic representations of the point group $G$?  In the case where $G=\mathrm{SO}(3)$ in $d=3$ spatial dimensions, some of us have addressed this question in detail in the recent paper \cite{fracton_magneto}.  Here, we provide a more abstract and general treatment of the problem, with a particular focus on discrete groups $G$ where exotic structures can arise.  As part of our discussion, we will consider the possibility of unconventional theories with broken time-reversal symmetry, and discuss whether hydrodynamics might be unstable to fluctuations (a la the flow of the Navier-Stokes equations in $d=1$ to the KPZ universality class \cite{spohn_nonlinear_2014}). We will review the effective field theory framework we use to answer these questions in Section \ref{sec:eft}, and describe the resulting hydrodynamics (usually diffusive) in Section \ref{sec:exoticcharges}, paying particular attention to the exotic conservation laws that can arise. 

The most interesting such theory which we have found, and which forms the basis of the second part of this paper, is a priori very simple: a theory in two spatial dimensions with triangular ($\mathrm{D}_3$ to physicists; $\mathrm{D}_6$ to mathematicians) point group, with a vector conserved charge and a generic current.  In this paper we will refer to $\mathrm{D}_6$ henceforth as the symmetry group. One can think of this intuitively as keeping only the momentum of the usual Navier-Stokes equations as a genuinely conserved quantity.  Within the \emph{isotropic} Navier-Stokes equations, one can easily see that the only dynamics which can result from such a truncation is the diffusive (viscous) relaxation of the vector charge.  With triangular symmetry, there is a naive possibility of finding a \emph{ballistic contribution} to this viscous mode. Yet recent work has found that such a ballistic contribution does not exist, either because it violated the KMS-invariance of the geometric action (in the case where the vector conserved charge is momentum) \cite{Huang_2022}, or because it is not compatible with kinetic theory of liquids with anisotropic kinetic energy \cite{Friedman_triangular}.  This raised the intriguing possibility that there may truly be constraints on hydrodynamics, arising from fundamental statistical mechanics, that are wholly invisible within the canonical Landau paradigm.

In this paper, we begin to resolve this puzzle: the terms described above are forbidden in a theory with a vector U(1) conservation law the absence of a \emph{triangular chiral anomaly}.  In conventional physical settings, such chiral anomalies can only arise in odd spatial dimensions $d$.  This is not due to a fundamental physics reason, but rather a group theoretic one: the only tensor which can be included in the anomalous terms in the hydrodynamic equations is the spacetime Levi-Civita tensor, contracted into the U(1) Maxwell tensor $F_{\mu\nu}$; hence $d$ must be odd.  In the triangular theory, it will turn out there is a spatial third-rank tensor which can play a similar role.  We discuss this anomaly in Section \ref{sec:anomalies} and in further detail in Appendix \ref{app:A}.   In the case where our vector conserved charge is instead momentum, this may suggest an unusual kind of anisotropic gravitaitonal anomaly \cite{Alvarez-Gaume:1983ihn}.

One might think that this anomaly is a curious quantum mechanical effect, but in fact, it can arise in a strictly classical system!  In Section \ref{sec:numerics}, we present extensive Markov chain simulations of a time-reversal- and inversion-breaking theory on a triangular lattice in two-dimensions, with a vector conserved charge.  We find unambiguous signatures of the anomalous hydrodynamics in this wholly classical setting.  Our model can ultimately be understood as an interesting generalization of how a certain biased random walk can realize the usual chiral anomaly in 1+1 dimensional theories.

\section{Review of effective field theory framework} \label{sec:eft}
 In this section we will review the effective field theory framework proposed in \cite{Guo:2022ixk}.  The effective field theory describes ``non-thermal" fluctuating systems with local, ergodic dynamics.  Here the phrase ``non-thermal" refers to the fact that there is no conserved energy and thus no temperature.  Nevertheless, we will posit the existence of a many-body stationary probability distribution for the stochastic dynamics which will lead to emergent notions of thermodynamics.
 
 Let $\rho(x,t)$ denote the density of a scalar conserved charge in $d$ spatial dimensions.  We will write down an action involving both $\rho$ and a conjugate ``noise field" $\pi$, of the form\footnote{In the formalism of \cite{crossley_effective_2017}, $\rho$ would be related to the $r$-field $\partial_t \phi_r$, and $\pi$ the $a$-field $\phi_a$, on the Schwinger-Keldysh contour.}\begin{eqnarray}
        S = \int \mathrm{d}t \mathrm{d}x \; \left[\pi \partial_t \rho - H(\pi,\rho)\right].
\end{eqnarray}
Here $H$ is a function to be determined, but we demand it to have no $\pi$-independent terms (this is roughly related to the desire that $\rho$ undergoes a stochastic process with normalized probability distribution): \begin{eqnarray}
        H(0,\rho)=0.
\end{eqnarray}The $\pi$ equation of motion gives us $\partial_t\rho = \cdots$, so the right hand side will encode the equations of motion for $\rho$.

Suppose that the many-body probability distribution is \begin{eqnarray}
         P_{\mathrm{eq}}[\rho] = \mathrm{e}^{-\Phi[\rho]}.
 \end{eqnarray}
 Defining a conjugate chemical potential \begin{eqnarray}
         \mu(x) = \frac{\delta \Phi}{\delta \rho(x)},
 \end{eqnarray}
 it was shown in \cite{Guo:2022ixk} that (in the weak noise or linear response limit, either of which is sufficient for our purposes here), time-reversal corresponds to the transformations $t\rightarrow -t$ and  \begin{eqnarray}\label{trs}
         \pi \rightarrow -\pi + \mathrm{i}\mu,
 \end{eqnarray}
 assuming (as we do here) that $\rho$ is even under time-reversal.  Moreover, in order to demand that charge is conserved:\begin{eqnarray}
      0=  \frac{\mathrm{d}}{\mathrm{d}t} \int \mathrm{d}x \; \rho,
\end{eqnarray}
we demand that (the integral of) $H$ is invariant under \begin{eqnarray}
         \pi \rightarrow \pi + c(t)
 \end{eqnarray}
 for arbitrary $x$-independent function of time $c(t)$.   Spatial parity is straightforward ($x\rightarrow -x$) and does nothing interesting to either $\rho$ or $\pi$.  Lastly, the assumption that statistical fluctuations are bounded forces \begin{eqnarray} \label{eq:imH}
\mathrm{Im}(H) \le 0.        
\end{eqnarray}

With these constraints, in one spatial dimension ($d=1$), the leading order terms $H$ that we can write down is
\begin{align} \label{Hc2}
    H = A(\rho)\partial_x \pi - \mathrm{i} \sigma(\rho)\,\partial_x\pi\,\partial_x(\pi-\mathrm{i} \mu) + \cdots.
\end{align}
where ${A(\mu)}$ and ${\sigma(\mu)}$ are functions of ${\mu}$ with no derivatives. Moreover, ${A(\rho)}=0$ if the system has P (parity) and/or T (time-reversal) symmetry.  The $\sigma$  term is compatible with both P and T symmetry, and is the minimal action for hydrodynamics for a single conserved charge.  Note that (\ref{eq:imH}) implies $\sigma \ge 0$, which is positivity of the conductivity and diffusion constant.  Indeed, the noise-free equation of motion for $\rho$ is found by varying $S$ with respect to $\pi$, and then setting $\pi \rightarrow 0$: \begin{eqnarray}
        \chi \partial_t \mu - \partial_x \left(\sigma \partial_x \mu \right) = 0.
\end{eqnarray}
This is the form of a standard continuity equation where the charge current obeys Fick's Law of diffusion.  Here $\chi = \partial \rho /\partial \mu$ is the charge susceptibility, and is a constant within linear response.

In this theory, the relative scaling dimension between time and space (dynamical critical exponent $z$) is given by $z=2$. Since the term ${\pi \partial_t \rho}$ has to be marginal, the scaling dimensions of ${\rho}$ and ${\pi}$ satisfy $[\rho] + [\pi] = d$. From (\ref{trs}) and ${\mu \sim \rho}$, we get \begin{eqnarray}
        [\rho] = [\pi] = \frac{d}{2}. \label{eq:scalingdims}
\end{eqnarray}.

If the system has PT symmetry (but not P or T separately), and the system is defined on a spatial circle with periodic boundary conditions, there is essentially no constraint on $A$.   After all \begin{eqnarray}
        \int \mathrm{d}x \; A(\mu) \partial_x \pi \rightarrow \int \mathrm{d}x \;  A(\mu) (-\partial_x) (-\pi + \mathrm{i}\mu) =\int \mathrm{d}x \;  A(x)\partial_x \pi + \mathrm{i} \int \mathrm{d}x \; A(\mu)\partial_x \mu.
\end{eqnarray} 
The last term is a total derivative and vanishes with periodic boundary conditions, meaning that the integral of $H$ is indeed invariant.  If the ${A(\rho)}$ term is nonzero, this term becomes the leading dissipationless term and can lead to instability.  Note that although $A$ can contribute a term to the equation of motion \begin{eqnarray}
       \chi \partial_t \mu  - \partial_x \left(A \partial_x \mu + \sigma \partial_x \mu \right) = 0. 
\end{eqnarray}within linear response (where $\mu$ and $\rho$ are proportional), we do not consider this to modify the dynamical critical exponent: it is more important to maintain $z=2$ so that fluctuations are not treated as irrelevant (it is better to instead imagine ``boosting" to a new reference frame and undoing the linear-in-$A$ term).

Now consider the leading nonlinear contribution from ${A(\rho)}$ to the current: \begin{eqnarray}
        J_x = \cdots + A_2 \mu^2 + \cdots.
\end{eqnarray}The scaling dimension of $[\mu]^2 = d$, which is smaller than or equal to that for the dissipative term $[\partial_x \mu] = 1+\frac{d}{2}$ when $d\le 2$.  In $d=1$, the nonlinearity is relevant and drives an instability of the hydrodynamic theory.  This is the instability of the Burgers equation, well-established in one dimension: it is well-known that the endpoint of this instability is the Kardar-Parisi-Zhang universality class \cite{spohn_nonlinear_2014}, which has anomalous exponent $z=3/2$.

\section{Theories with exotic conserved charges}
\label{sec:exoticcharges}
We now extend the discussion of the previous section to more general theories where the conserved charge $\rho_a$ transforms in a non-trivial irreducible representation of a spatial point group $G$ associated to the rotational symmetry.
\subsection{General framework} \label{sec:generalframework}

Suppose the microscopic dynamics are invariant under space group $G$, and suppose there is a conserved charge $\rho_a$ and current $J_\alpha$ which transform in possibly non-trivial  representations of $G$. For simplicity, we take $\rho_a$ to transform as an irreducible representation $\mathcal{R}_a$; if it were reducible, we could equivalently consider each irrep to be a separately conserved quantity. We allow $J_\alpha$ to be more general and transform in a possibly reducible representation $\bigoplus_{i} \mathcal{R}_{\alpha_i}$. A general (non-multipolar) conservation law has the form 
\begin{equation} \label{eq:generalhydro}
    \partial_t \rho_a + \partial_i \Gamma_{ia\alpha} J_\alpha = 0,
\end{equation}
where $\Gamma_{ia\alpha}$ is a set of generalized Clebsch-Gordan coefficients. The $\Gamma_{ia\alpha}$ are nonzero when $\mathcal{R}_a$ appears in the irrep decomposition of $(\bigoplus_k \mathcal{R}_{\alpha_k}) \otimes \mathcal{V}_i$ ($\mathcal{V}_i$ denotes the $d$-dimensional vector representation in which the derivative lies). For $G = \mathrm{SO}(3)$ and $\rho_i$ transforming as a vector, and different choices of $J_\alpha$, we recover known aspects of hydrodynamics with vector conserved currents, as was discussed at some length in a recent paper \cite{fracton_magneto}.

Eq. \eqref{eq:generalhydro} leads to (possibly infinitely many) conserved quantities. To find them, consider the quantity \begin{eqnarray}
        \mathcal{Q}[f_a] := \int\mathrm{d}^dx\;  f_a \rho_a,
\end{eqnarray} where $f_a$ are arbitrary functions of space. This quantity being conserved means its time derivative vanishes; imposing this as a condition (and assuming periodic boundary conditions, or that $\rho_a$ vanishes at infinity) gives 
\begin{equation}
    0 = \frac{\mathrm{d}}{\mathrm{d}t} \int f_a \rho_a = -\int \Gamma_{i a \alpha} f_a \partial_i J_\alpha = \int J_\alpha \Gamma_{i a \alpha} \partial_i f_a.
\end{equation}
We therefore find that the quantity $\mathcal{Q}[f_a]$ are conserved when 
\begin{equation} \label{eq:fcondition}
        \Gamma_{i a \alpha} \partial_i f_a = 0.
\end{equation}
 When $\rho_a$ are the only conserved charges, and $J_\alpha $ lies in an reducible representation as well, in general the lowest order term in the gradient expansion is 
\begin{equation} \label{eq:constitutivereln}
    J_\alpha = \sum_{\alpha_k} -D_{\alpha_k} \Gamma_{ib\alpha_k} \partial_i \rho_b
\end{equation}
which leads to the generalized diffusion equation 
\begin{equation}
    \partial_t \rho_a - \sum_{\alpha_k} D_{\alpha_k} \Gamma_{ia\alpha_k} \Gamma_{jb\alpha_k} \partial_i \partial_j \rho_b = 0.
\end{equation}
Note here that each representation $\mathcal{R}_{\alpha_k}$ would in general get its own diffusion constant $D_{\alpha_k}$.

We can reformulate the above discussion in terms of the hydrodynamic effective field theory of Section \ref{sec:eft}. We generalize slightly the construction of the previous section to allow the density $\rho_a$ and conjugate field $\pi_a$ to transform nontrivially under the space group $G$. The action then takes the form


\begin{equation}
    S = \int \mathrm{d}t \mathrm{d}x \; \left[  \pi_a \partial_t \rho_a - H(\pi_a, \rho_a)\right]
\end{equation}
where $H(\pi_a, \rho_a)$ obeys analogous constraints as in Sec. \ref{sec:eft}. In particular, the existence of a steady-state mandates 
\begin{equation} \label{eq:steadystate}
    \int \mathrm{d}x \; H(0, \rho_a) = \int \mathrm{d}x \; H(\mathrm{i}\mu_a , \rho_a)  = 0. 
\end{equation}
To encode the conservation law \eqref{eq:generalhydro}, we require that $\pi_a$ only appear in $H(\pi_a, \rho_a)$ via the combination $\Gamma_{i a \alpha} \partial_i \pi_a$. This implies that the Hamiltonian is invariant under the transformation $\pi_a \to \pi_a + f_a(x,t)$, where $f_a (x,t)$ satisfies \eqref{eq:fcondition}. And as before, well-posedness of statistical fluctuations imposes the condition \eqref{eq:imH}. Given these constraints, the most general Hamiltonian we can write is 
\begin{equation} \label{eq:dissipativehamiltonian}
    H(\pi_a, \rho_a) =  - \mathrm{i} \sum_{\alpha_k} \sigma_{\alpha_k} (\rho) \Gamma_{i a \alpha_k} \partial_i \pi_a \Gamma_{j b \alpha_k} \partial_j (\pi_b - \mathrm{i} \mu_b) + \ldots
\end{equation}
where $\sigma(\rho_a)$ is a function of $\rho_a$ with no derivatives. This action leads to the equation of motion 
\begin{equation}
    \partial_t \rho_a - \sum_{\alpha_k} \Gamma_{i a \alpha_k} \partial_i (\sigma_{\alpha_k} (\rho_a) \Gamma_{j b \alpha_k} \partial_j \mu_b) = 0,
\end{equation}
which can be identified with the continuity equation \eqref{eq:generalhydro} and constitutive relation \eqref{eq:constitutivereln} at linear order after identifying 
\begin{equation}
D_{\alpha_k} = \frac{\sigma_{\alpha_k}(\bar{\rho}_a)}{\chi}
\end{equation}
where $\bar{\rho}_a$ is the average charge density and $\chi$ is the susceptibility defined as $\chi \delta_{ab} = \frac{\partial \rho_a}{\partial \mu_b}$. 

One can also consider the possibility of dissipationless terms in the constitutive relation \eqref{eq:constitutivereln}. In particular, this can happen when $\mathcal{R}_a$ appears in the irrep decomposition $\bigoplus_{i} \mathcal{R}_{\alpha_i}$. Then a term such as 
\begin{equation} \label{eq:dissipationlessJ}
    J_\alpha \supset v \rho_a \delta_{\alpha a}
\end{equation}
is allowed on group theoretic grounds. Here the Kronecker delta indicates an inclusion of the $\mathcal{R}_a$ subrepresentation into $\bigoplus_i \mathcal{R}_{\alpha_i}$ However, it is a priori unclear whether such a term is thermodynamically consistent. This is where the effective field theory formalism proves especially useful, as it provides a systematic method of determining whether such terms are permitted. This term in the constitutive relation corresponds to a term 
\begin{equation} \label{eq:dissipationlessH}
    H \supset -\alpha \mu_a \Gamma_{iab} \partial_i \pi_b
\end{equation}
in the Hamiltonian. This manifestly satisfies invariance under $\pi_a \to \pi_a + f_a(x,t)$ as well as well-posedness of statistical fluctuations, but is not in general consistent with the existence of a steady state as \eqref{eq:steadystate} is not obeyed. We note, however, that if $\Gamma_{iab}$ is symmetric with respect to $a$ and $b$, then it is possible for the term \eqref{eq:dissipationlessH} to be a total derivative upon substituting $\pi_a = \mathrm{i} \mu_a$. Explicitly, 
\begin{equation}
    -\int \mathrm{d}x \; \alpha \mu_a \Gamma_{iab} \partial_i (\mathrm{i} \mu_b) = -\mathrm{i} \int \mathrm{d}x \;  \alpha \mu_a \Gamma_{iab} \partial_i \mu_b = -\mathrm{i} \int \mathrm{d}x \; \partial_i \left(\frac{1}{2} \alpha \Gamma_{iab} \mu_a \mu_b\right) = 0,
\end{equation}
so the term \eqref{eq:dissipationlessH} satisfies the condition \eqref{eq:steadystate} if $\Gamma_{iab}$ is symmetric in its last two indices and $\alpha$ is constant. The velocity in \eqref{eq:dissipationlessJ} is related to $\alpha$ and the susceptibility by $v = \alpha / \chi$.


\subsection{Hydrodynamics with triangular symmetry} \label{sec:hydrowithtrianglesymmetry}
We now specialize to the hydrodynamics of a two-dimensional system which is invariant under the point group an equilateral triangle, $\mathrm{D}_6$. Let us first recall some useful properties of $\mathrm{D}_6$. $\mathrm{D}_6$ has three irreducible representations: the trivial representation $\mathbf{1}$, the sign representation $\mathbf{1'}$, and the two dimensional representation $\mathbf{2}$. The two dimensional representation is the vector representation, which is acted on by $\mathrm{D}_6$ via $2 \times 2$ matrices viewed as a subgroup of $\mathrm{O}(2)$. What is unique about this restriction is that the traceless symmetric tensors (which form a two-dimensional ``spin 2" irreducible representation of $\mathrm{O}(2)$) \emph{also} transform as the vector representation $\mathbf{2}$ under $\mathrm{D}_6$. The multiplication table of irreps of $D_6$ is as follows:
\begin{center}
\begin{tabular}{c| c c c}
                & $\mathbf{1}$ & $\mathbf{1'}$ & $\mathbf{2}$ \\
                \hline
                $\mathbf{1}$ & $\mathbf{1}$ & $\mathbf{1'}$ & $\mathbf{2}$ \\
                $\mathbf{1'}$ & $\mathbf{1'}$ & $\mathbf{1}$ & $\mathbf{2}$ \\
                $\mathbf{2}$ & $\mathbf{2}$ & $\mathbf{2}$ & $\mathbf{1} \oplus \mathbf{1'} \oplus \mathbf{2}$
\end{tabular}
\end{center}
The independent invariant tensors of $\mathrm{D}_6$ are $\delta_{ij}$ and $\lambda_{ijk}$. The first is inherited from the two-dimensional rotation group $\mathrm{O}(2)$, while $\lambda_{ijk}$ is intrinsic to $\mathrm{D}_6$. The components of $\lambda_{ijk}$ are 
\begin{equation} \label{eq:lambda}
        (\lambda_1)_{ij} = -\sigma^z_{ij} = \begin{pmatrix} -1 & 0 \\ 0 & 1 \end{pmatrix}, \;\; (\lambda_2)_{ij} = \sigma^x_{ij} = \begin{pmatrix} 0 & 1 \\ 1 & 0 \end{pmatrix}.
\end{equation}
One can check that $\lambda_{ijk}$ is completely symmetric, and its trace over any two indices is zero. Intuitively, $\lambda_{ijk}$ can be seen as converting between vector and traceless-symmetric tensor interpretations of $\mathbf{2}$. 

We will be interested in hydrodynamics where the charge is a vector. In this case, the conservation law reads
\begin{equation}
    \partial_t \rho_i + \partial_j J_{ij} = 0.
\end{equation}
In general, $J_{ij}$ can be decomposed into the trace, antisymmetric, and traceless symmetric parts, which correspond to the $\mathbf{1}$, $\mathbf{1'}$, and $\mathbf{2}$ irreps of $D_6$, respectively. For generic $J_{ij}$ containing all three irreps, the only conserved quantities are $\int \rho_i$. 

We can use this as a starting point to build the hydrodynamic effective field theory described in Sec. \ref{sec:generalframework}. The action is 
\begin{equation}
    S = \int \mathrm{d}t \mathrm{d}x \mathrm{d}y \; \left[ \pi_i \partial_t \rho_i - H(\pi_i, \rho_i)\right]
\end{equation}
and we take $H(\pi_i, \rho_i)$ to be 
\begin{equation}
\begin{aligned}
    H(\pi_i, \rho_i) = &-\mathrm{i} \sigma_1 \bigg(\partial_i \pi_j + \partial_j \pi_i - \delta_{ij} (\partial \cdot \pi) \bigg) \bigg(\partial_i (\pi_j - \mathrm{i} \mu_j) + \partial_j (\pi_i - \mathrm{i} \mu_i) - \delta_{ij} \partial \cdot (\pi - \mathrm{i}\mu) \bigg) \\
            &-\mathrm{i} \sigma_2 (\partial \cdot \pi) \bigg(\partial \cdot (\pi - \mathrm{i}\mu) \bigg) 
            -\mathrm{i} \sigma_3 (\partial_j \pi_i - \partial_i \pi_j) \bigg( \partial_j (\pi_i - \mathrm{i} \mu_i) - \partial_i (\pi_j - \mathrm{i} \mu_j) \bigg) \\
            &- \alpha \mu_i \lambda_{ijk} \partial_j \pi_k .
\end{aligned}
\end{equation}
The first three terms in $H(\pi_i, \rho_i)$ are the terms of \eqref{eq:dissipativehamiltonian} for each irrep of the triangular point group. The last term is a dissipationless contribution which is possible because $\lambda_{ijk}$ is symmetric in all of its indices. After making the identifications $D_1 = 2 \sigma_1/\chi$, $D_2 = \sigma_2 /\chi$, $D_3 = 2 \sigma_3/\chi$ and $v= \alpha / \chi$, the Hamiltonian terms correspond to the constitutive relation 
\begin{equation} \label{eq:triangleconstitutive}
    J_{ij} = v \lambda_{ijk} \rho_k - D_1   \underbrace{\big(\partial_i \rho_j + \partial_j \rho_i - \delta_{ij} (\partial \cdot \rho)\big) }_\mathbf{2} - D_2 \underbrace{(\partial \cdot \rho) \delta_{ij}}_{\mathbf{1}} - D_3 \underbrace{ (\partial_j \rho_i - \partial_i \rho_j)}_{\mathbf{1'}} .
\end{equation}
This leads to the equation of motion
\begin{equation} \label{eq:trianglehydroeom}
    \partial_t \rho_i + v \lambda_{ijk} \partial_j \rho_k - (D_1 + D_3) \partial^2 \rho_i - (D_2 - D_3) \partial_i (\partial \cdot \rho) = 0.
\end{equation}
From the effective action we can show that the two-point correlation functions $C_{ij}(x,t) = \langle \rho_i(x,t) \rho_j(0,0) \rangle$ are the Green's functions for the equations of motion \eqref{eq:trianglehydroeom}. Let us consider a simplified situation where $D_2$ and $D_3$ are equal so the last term of \eqref{eq:trianglehydroeom} vanishes, and let $D = D_1 + D_3$. We solve for $C_{ij}$ in Fourier space, where \eqref{eq:trianglehydroeom} takes the form
\begin{equation} \label{eq:twopointdiffeqfourier}
    (-\mathrm{i} \omega \delta_{il} - i v k_k  \lambda_{ikl} + D k^2 \delta_{il} ) C_{lj} = 0.
\end{equation}
This can be interpreted as an eigenvalue equation for the matrix $-v k_k \lambda_{kil} - i D k^2 \delta_{il}$ with eigenvalue $\omega$. The eigenvalues are  $\omega = -i D k^2 \pm v k$ with corresponding eigenvectors 
\begin{equation}
    u^+ = \begin{pmatrix} \sin{\frac{\theta}{2}} \\ \cos{\frac{\theta}{2}} \end{pmatrix}, \;\; u^- =  \begin{pmatrix} -\cos{\frac{\theta}{2}} \\ \sin{\frac{\theta}{2}} \end{pmatrix}
\end{equation}
where $\theta$ is the angle between $\vec{k}$ and the $x$-axis.  The full solution to \eqref{eq:twopointdiffeqfourier} in $k$-space is 
\begin{equation}
    C_{ij}(\vec{k},t) = c^+_j u^+_i \mathrm{e}^{ - \mathrm{i} \omega^+(k) t} + c^-_j u^-_i \mathrm{e}^{- \mathrm{i}\omega^-(k) t}.
\end{equation}
The initial condition in $k$-space is $C_{ij}(\vec{k},t=0) = \delta_{ij}$, which sets $c^+_j = u^+_j$, $c^-_j = u^-_j$. Therefore we have
\begin{equation}\label{eq:3C}
    C_{ij}(\vec{k},t) = \mathrm{e}^{-D k^2 t}  \begin{pmatrix} \cos{v k t} - \mathrm{i} \cos{\theta} \sin{v k t} & \mathrm{i} \sin{\theta} \sin{v k t} \\
    \mathrm{i} \sin{\theta} \sin{v k t} & \cos{v k t} + \mathrm{i} \cos{\theta} \sin{v k t} \end{pmatrix}.
\end{equation}
We will use this Green's function to diagnose the presence of T-broken hydrodynamics in our numerical simulations in Section \ref{sec:numerics}.

Lastly, let us remark on the hydrodynamic stability of this theory.   Assuming locality, the leading order expression for $\Phi$ (defined in Section \ref{sec:generalframework}) is \begin{eqnarray}
        \Phi = F(\rho_i \rho_i, \lambda_{ijk}\rho_i\rho_j\rho_k). \label{eq:phiF}
\end{eqnarray}
Hence, the leading order terms in $\mu_i$ are \begin{eqnarray}
        \mu_i = b_1 \rho_i + b_2 \lambda_{ijk}\rho_j\rho_k + b_3 \rho_i \rho_j\rho_j + \cdots.
\end{eqnarray}
where $b_{1,2,3}$ denote $\rho$-independent constants.  The power counting for $[\pi]$ and $[\rho_i]$ follows along the same lines as (\ref{eq:scalingdims}).  If $b_2 \ne 0$ (note that then $b_3 \ne 0$ is required for stability purposes), then there are marginal nonlinearities in this theory.  While we do not know the ultimate impact of these nonlinearities on the nature of the hydrodynamic fixed point, it is likely that they are not so important in practice: even in the two-dimensional Navier-Stokes equations where such a nonlinearity is marginally \emph{relevant}, its effects are rather weak in practice (e.g. one uses two-dimensional hydrodynamics routinely to model experiments!).

\subsection{Holomorphic conserved charges} \label{sec:holomorphicconservation}
It is interesting to examine the special case where the current $J_{ij} = \lambda_{ijk} J_k$ is restricted to live in the vector representation. In this case, the conservation law reads 
\begin{equation} \label{eq:holomorphicconservation}
    \partial_t \rho_i + \lambda_{ijk} \partial_j J_k = 0. 
\end{equation}
Applying \eqref{eq:fcondition}, the conserved quantities $\mathcal{Q}[f]$ satisfy \begin{eqnarray}
        \lambda_{ijk} \partial_j f_k = 0.
\end{eqnarray} Expanded out using \eqref{eq:lambda}, the $f_i$ obey 
\begin{equation}
\begin{aligned}
    -\partial_x f_x + \partial_y f_y &= 0 \\
    \partial_x f_y + \partial_y f_x &= 0 
\end{aligned}
\end{equation}
which are the Cauchy-Riemann equations for $f_i$. It follows that any holomorphic function $f(z)$ yields a corresponding conserved quantity. We can identify an infinite generating set of conserved quantities as coming from $f_n(z) = z^n$ and $\tilde{f}_n = \mathrm{i} z^n$ for $n$ a nonnegative integer. We will refer to these conserved quantities as holomorphic moments.

While the existence of an infinite family of conserved quantities may at first seem fine-tuned, these can in fact emerge naturally as "quasiconserved" quantities in the sense of \cite{quasiconservation}. Suppose the microscopic dynamics enforced the conservation of $D = \int \vec{r} \cdot \vec{\rho}$ and $L = \int \vec{r} \times \vec{\rho}$. These correspond to the holomorphic functions $f(z) = z$ and $f(z) = \mathrm{i}z$, respectively. In order for $D$ and $L$ to be conserved, the continuity equation must take the form of \eqref{eq:holomorphicconservation} within linearized hydrodynamics at leading order in the derivative expansion; as a result, the holomorphic moments emerge as an infinte tower of conserved quantities. However, this is only true at leading order in linearized hydrodynamics; higher order terms in the hydrodynamic expansion (as well as nonlinear terms) may cause the moments to decay. As such, they decay subdiffusively, in contrast to what would be expected from the form of the continuity equation. Hence the higher holomorphic moments would be "quasiconserved" since they decay parametrically slowly. The physics is similar to the situation discussed in \cite{quasiconservation} where the existence of only finitely many harmonic functions in two dimensions can also lead to an infinite family of such quasiconserved quantities in fracton hydrodynamics.  

\subsection{Other dihedral groups}
The picture outlined above generalizes straightforwardly to the case of odd dihedral groups: see e.g. \cite{caleb}. In general, for dihedral groups $\mathrm{D}_{2n}$ with $n$ odd, the two-dimensional spin-$k$ irreps of $\mathrm{O}(2)$ for $k = 1, 2, \ldots, \frac{n-1}{2}$ descend to two-dimensional irreps of $\mathrm{D}_{2n}$. The two one-dimensional irreps of $\mathrm{O}(2)$ similarly descend to $\mathrm{D}_{2n}$.  We denote the spin-$k$ irreps as $\mathbf{2}_k$, and one-dimensional irreps as $\mathbf{1}$ and $\mathbf{1'}$.  The ordinary vector representation is $\mathbf{2}_1$, and spin-$k$ representations can be identified with completely traceless-symmetric tensors with $k$ indices. The group contains a completely traceless-symmetric invariant tensor $\lambda_I$, with $I = i_1 \ldots i_n$ a multi-index tensor. The construction of this invariant tensor parallels that of $\lambda_{ijk}$ in \eqref{eq:lambdaconstruction1} and \eqref{eq:lambdaconstruction2}. The role of this invariant tensor is to identify the spin-$k$ and spin-$|k-\mathbb{N}n|$ representations of $SO(2)$ in $\mathrm{D}_{2n}$. The multiplication table for the tensor product of irreps descends from that of $\mathrm{O}(2)$ up to the identification provided by $\lambda_I$ \cite{caleb}. 

For concreteness we can take $n=5$ as an example. For a charge $\rho_{ij}$ which is traceless-symmetric transforming in the $\mathbf{2}_2$ representation of $\mathrm{D}_{10}$, the general conservation law takes the form
\begin{equation}
    \partial_t \rho_{ij} + \partial_k J_{ijk} = 0.
\end{equation}
The presence of the invariant tensor allows for a term 
\begin{equation}
    J_{ijk} = v \lambda_{ijklm} \rho_{lm} + \ldots
\end{equation}
in the constitutive relation. Because $\lambda_{ijklm}$ is completely symmetric, this term is allowed effective field theory formalism of Sec. \ref{sec:generalframework}. Identical considerations apply to $\rho_I$ transforming in the $\mathbf{2}_\frac{n-1}{2}$ irrep of $\mathrm{D}_{2n}$.  

Something which differs between this theory and the $\mathrm{D}_6$-invariant theory discussed previously is that we cannot generalize (\ref{eq:phiF}): there is no way to contract three copies of $\rho_{lm}$ with $\lambda_{ijklm}$ and $\delta_{ij}$.  Therefore, there is no $\rho^2$ term in $\mu$; we conclude that there are no marginal nonlinear operators that can be added to the hydrodynamic action.  Hence the hydrodynamic theory identified above is a strictly stable fixed point for the $n=5,7,\ldots$ theories.

\section{Anomalies} \label{sec:anomalies}
In this section, we will now explain that the drift term ($v$) captured in (\ref{eq:triangleconstitutive}) is in fact the consequence of a chiral anomaly.  Our discussion here will be somewhat brief, as we only wish to explain the effect from the perspective of classical physics and hydrodynamics.  A discussion of a quantum mechanical theory with this anomaly, which closely mirrors the recent paper \cite{subsysteminflow}, is contained in Appendix \ref{app:A}. 
\subsection{Warm-up: biased random walk}
As a warm-up example, we first review the hydrodynamics of a biased diffusion process in one dimension, which arises when we only have PT symmetry.  It is known in isotropic fluids how  a hydrodynamic effective theory can capture a U(1) chiral anomaly \cite{luca}, using a more sophisticated geometric construction than what was described in Section \ref{sec:eft}.  But one can also understand this anomalous fluid dynamics, already at the ideal hydrodynamic level, by a simple Hamiltonian system for density $\rho(x)$ with a modified Poisson bracket:
\begin{equation} \label{eq:chiralbosonpoisson}
    \{ \rho(x), \rho(y) \} = \partial_x \delta(x-y).
\end{equation}
The Poisson bracket satisfies anticommutativity since 
\begin{equation}
\{\rho(y), \rho(x)\} =  \partial_y \delta(y-x) = - \partial_x \delta(y-x) = - \partial_x \delta(x-y) = -\{\rho(x), \rho(y) \}.
\end{equation} 

Now consider Hamiltonian 
\begin{equation}
    H = \int \mathrm{d}x \; \frac{1}{2} v \rho(x)^2.
\end{equation}
The Hamilton equations of motions are 
\begin{equation}
    \dot{\rho}(x) = \{\rho(x), H\} = v \partial_x \rho(x)
\end{equation}
which is precisely the hydrodynamic equation of motion for the biased random walk at the non-dissipative level. The Poisson bracket \eqref{eq:chiralbosonpoisson}, when quantized, appears in the commutation relations of the chiral boson \cite{Wen:1995qn}, which has an anomaly associated to its $\mathrm{U}(1)$ symmetry \cite{SONNENSCHEIN1988752}.

\subsection{Triangle fluid}
The discussion for the time-reversal-breaking fluid with triangular point group proceeds similarly. Like the biased random walk, the hydrodynamics of this theory is anomalous.  We can see this anomaly arise at the classical level via Hamiltonian dynamics: a quantization of this theory is found in Appendix \ref{app:A}. 

Here we have two conserved quantities $\rho_x$ and $\rho_y$. The Poisson brackets for these fields are 
\begin{equation}
    \{ \rho_i(x), \rho_j(y) \} = \frac{1}{a} \lambda_{ijk} \partial_k \delta^2(x-y) \label{eq:4P}
\end{equation}
where $\lambda_{ijk}$ is the invariant $\mathrm{D}_6$ tensor described earlier. Owing to the symmetry of $\lambda_{ijk}$, the Poisson bracket is antisymmetric in the same way as in the biased random walk case. A new feature in this case is the existence of a length scale $a$, which is needed on dimensional grounds. The Hamiltonian is 
\begin{equation}
    H = \int \mathrm{d}^2 x \; \frac{1}{2} a v (\rho_x^2 + \rho_y^2). \label{eq:4H}
\end{equation}
Note that the length scale $a$ appears explicitly in the Hamiltonian as well. The Hamiltonian equations of motion are 
\begin{equation}
    \dot{\rho}_i = v \lambda_{ijk} \partial_j \rho_k
\end{equation}
which reproduce the equations of motion \eqref{eq:trianglehydroeom} at the non-dissipative level. The structural similarities with the biased random walk in the previous section suggest that the physics is controlled by an anomaly similar to the that of the chiral boson in one dimension. Indeed, in Appendix \ref{app:A} we propose and analyze a field theory exhibiting such an anomaly. 

An unusual feature of the theory is that there is a length scale $a$ which appears explicitly, both in (\ref{eq:4P}) and (\ref{eq:4H}).  At the classical level, we cannot say much more.  In quantum mechanics, analysis of the anomaly reveals that $a^{-1}$ is quantized in units of $L^{-1}$, where $L$ is the system size.  This would lead to an unusual kind UV-IR mixing, where the IR data enters into the UV commutator and Hamiltonian.  However, when interpreting an anomaly inflow problem whereby the $2+1$-dimensional anomaly is cancelled by a $3+1$-dimensional bulk action, the natural bulk action to write down suggests that $a^{-1}$ is a lattice spacing.   We leave a more detailed analysis of interpreting $a$ to future work.

\section{Markov chains}
\label{sec:numerics}

We now simulate hydrodynamics in systems with triangular symmetry using classical Markov chains, and observe compelling evidence for the anomalous hydrodynamics predicted above.

\subsection{Some useful facts}
Before describing the Markov chain, we briefly review a few useful textbook facts about the triangular lattice.  The lattice is built out of adjacent points connected by the unit vectors \begin{eqnarray}
        \mathbf{e}_1 = (1,0), \;\;\;\; \mathbf{e}_2 = \left(-\frac{1}{2},\frac{\sqrt{3}}{2}\right), \;\;\;\; \mathbf{e}_3 = \left(-\frac{1}{2},-\frac{\sqrt{3}}{2}\right). \label{eq:ebasis}
\end{eqnarray}
This orientation is depicted in Figure \ref{lattice}, and is quite useful due to the identity \begin{eqnarray}
        \mathbf{e}_1+\mathbf{e}_2+\mathbf{e}_3 = \mathbf{0}. \label{sumezero}
\end{eqnarray}
In our Markov chain, we will place a charge on each edge $e$ of the lattice.  The $x,y$-components of this vector charge are given by \begin{eqnarray}
        (q_x^e,q_y^e) = q_e \mathbf{e}_i,
\end{eqnarray}
where $\mathbf{e}_i$ is the orientation of that particular edge using the conventions of the figure.  Our Markov chains will only conserve the two quantities \begin{eqnarray}
        Q_{x,y} = \sum_{\text{edges }e} q_{x,y}^e.
\end{eqnarray}

There are two natural ways to find the tensor $\lambda_{ijk}$, which are natural to find using the isomorphism between the groups $\mathrm{D}_6$ and $\mathrm{S}_3$.  One finds that \begin{eqnarray} \label{eq:lambdaconstruction1}
        \lambda_{ijk} = \frac{2}{3}\sum_{\sigma \in \mathrm{S}_3} e_{i,\sigma(1)}e_{j,\sigma(2)}e_{k,\sigma(3)},
\end{eqnarray}
as well as \begin{eqnarray} \label{eq:lambdaconstruction2}
        \lambda_{ijk} = \frac{4}{3}\sum_{a=1}^3 e_{i,a}e_{j,a}e_{k,a}. 
\end{eqnarray}
These identities will give us useful clues as to where the anomalous hydrodynamics will arise in our simulations.

\subsection{Details of the Markov chains}
\begin{figure}[t]
\centering
\includegraphics[scale=0.48]{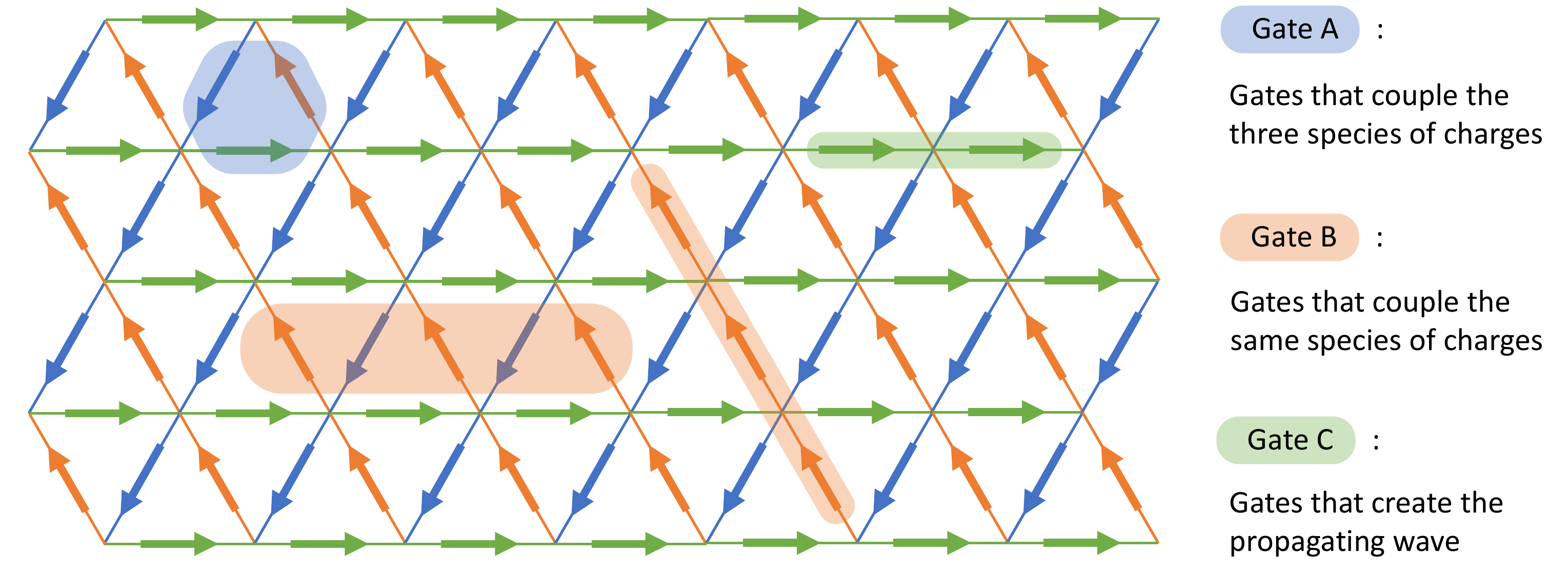}
\caption{The triangular lattice we used in our simulation. The directions and colors of the arrows represent the positive directions and different species of the charges. The blocks with different colors represent different kinds of gates and the charges on which they are acted. Note that although in this figure, gate B and gate C are acted on only one species of charges, they actually also act on all species of charges in a way that preserve the triangular symmetry.}
\label{lattice}
\end{figure}

We take a triangular lattice with periodic boundary conditions, and place a vector charge on each edge of the lattice, as shown in Figure \ref{lattice}. The allowed values of charges are ${q = 0, \pm 1, \cdots, \pm 4}$ (the precise value 4 here is not too important for what follows). 

The update rules of the Markov chain are best described pictorially, as shown in Figure \ref{lattice}.  We shortly provide more details.  First, we note the big picture: in each time interval, we randomly act with one of three different kinds of ``gates" (which replace charge configurations on nearby edges with other configurations, in a way that respects the conservation laws), labeled A/B/C.  The number of gates applied during each time interval is extensive: on an $L\times L$ lattice we apply $L^2$ gates per time step, drawn uniformly at random from the possibilities described above.

Gate A acts on a triangular plaquette of either orientation up or down.  Let $(q_1,q_2,q_3)$ denote the values of charges on each of the three edges of the lattice.  Then gate A will, with uniform probability, replace this configuration with another one of the form $(q_1+c,q_2+c,q_3+c)$, subject to the constraint that $|q_i+c|\le 4$.  This conserves both the $x$ and $y$ components of charge, as is seen straightforwardly using (\ref{sumezero}).

Gate B acts on three adjacent edges of the same orientation, and randomly replaces the charge configuration $(q_1,q_2,q_3)$ on these three edges with a different one, subject to the constraints that charges are at most $\pm 4$, and that $q_1+q_2+q_3$ is unchanged.

Gate C acts on two adjacent edges of the same orientation, and further oriented along the direction of the edge $\mathbf{e}_i$.  The update rule here is that whenever the absolute value of charge to the left (as defined by the edge at the tail of the orientation vector $\mathbf{e}_i$) is larger than the charge at the right, the two charges are swapped with probability $\frac{1}{8}(q_{\mathrm{left}} - q_{\mathrm{right}})$. 


Let us first prove that this Markov chain has the desired spacetime symmetry group.  It is obviously invariant under $120^\circ$ rotation.  Parity symmetry is a bit more subtle: the desired parity transformation turns out to be $(x,y)\rightarrow (x,-y)$, which (assuming the origin is a lattice point) effectively flips $\mathbf{e}_2$ and $\mathbf{e}_3$ -- again, the update rules are clearly invariant, as is importantly $\lambda_{ijk}$.  

In contrast, the transformation $(x,y)\rightarrow (-x,y)$ sends $\mathbf{e}_1\rightarrow -\mathbf{e}_1$, $\mathbf{e}_2\rightarrow -\mathbf{e}_3$, $\mathbf{e}_3 \rightarrow -\mathbf{e}_2$ -- this is not a symmetry of the theory.  The reason is that if $\mathbf{e}$ flips orientation, Gate C also ``reverses" and causes large charges to move left, rather than right. (In contrast, Gates A and B are unchanged, and the change in coordinates of any gates are not important since the update rules are discrete-translation invariant) We conclude based on this observation that without Gate C, this Markov chain is invariant under the full hexagonal symmetry group $\mathrm{D}_{12}$ and is time-reversal invariant, while when Gate C is included, the chain has manifest $\mathrm{D}_6$ invariance and is only invariant under time-reversal combined with inversion.  These are precisely the desired properties.


Next, we prove that the stationary distribution (up to conservation laws) of this Markov chain is uniform:  namely, all microstates are equally likely to be found.  This is a very useful property since we can easily sample from this distribution by simply initializing the chain in a uniformly random configuration:  we can then safely evaluate correlation functions of the form $\langle \rho_i(x,t)\rho_j(0,0)\rangle$ by simply running the chain for time $t$ and (after averaging over realizations, and space-time translations) looking at the average product of charges on two sites.   The proof proceeds by showing that for any microstate of the system, we are just as likely to transition into that state as to transition out of it. This reversibility holds even when we fix the location of Gate A or B, so clearly the chain as a whole is also time-reversal symmetric under Gates A and B.  Moreover, Gates A and B cannot admit a non-uniform (within fixed charge sector) stationary distribution: for each of these gates, the transition matrix (restricted to the sites the gate acts on) has a single non-null vector which is uniform.  Since using sufficiently many gates we can connect all microstates in the same charge sector to each other, we deduce that the unique many-body stationary distribution for Gates A and B is uniform.

Since Gate C breaks time-reversal symmetry, we need to consider the full microstate to prove that the transition rates in and out are equal.  Following \cite{Guo:2022ixk}, consider building a cycle (closed loop) on the lattice by starting with any edge $e$, and then appending the edge of the same orientation directly next to it (oriented along the appropriate $\mathbf{e}_i$).  Since the lattice is finite this process must terminate: call the resulting cycle $\Gamma = (e_1,e_2,\ldots)$.  Trivially, we have the following telescoping sum identity: \begin{eqnarray}
        \sum_i (q_{e_i} - q_{e_{i+1}}) = 0, \label{telescopingsum}
\end{eqnarray}
where (if the cycle has length $N$) we identify $e_1 = e_{N+1}$.  Observe that Gate C will flip charges with probability proportial to $q_{e_i} - q_{e_{i+1}}$ only when that difference is positive, with a rate proportional to that difference.  So the total transition rate out of this microstate (coming from Gate C acting along this cycle) is proportional to the sum of positive terms only in (\ref{telescopingsum}): \begin{eqnarray}
        P_{\mathrm{out}} = \frac{1}{8N} \sum_i (q_{e_i} - q_{e_{i+1}})\mathrm{\Theta}(q_{e_i} - q_{e_{i+1}}),
\end{eqnarray}
with $\mathrm{\Theta}$ the step function.
The prefactor arises from the fact that we are equally likely to act with Gate C anywhere along the cycle, and this calculation assumes that no other gates will act anywhere.  The transition rate into this microstate, on the other hand, arises from places where $q_{e_{i+1}}>q_{e_i}$, since whenever this happens, we could have (in the previous time step) have been in a state where those charges were flipped.  The total transition rate \emph{into} our microstate is then \begin{eqnarray}
        P_{\mathrm{in}} = \frac{1}{8N} \sum_i |q_{e_i} - q_{e_{i+1}}|\mathrm{\Theta}(q_{e_{i+1}}-q_{e_i}  ).
\end{eqnarray}
We clearly see that $P_{\mathrm{in}} = P_{\mathrm{out}}$, which ensures that the uniform distribution is stationary \cite{Guo:2022ixk,levin}.


\subsection{Numerical results}

\begin{figure}[t]
\centering
\includegraphics[scale=0.78]{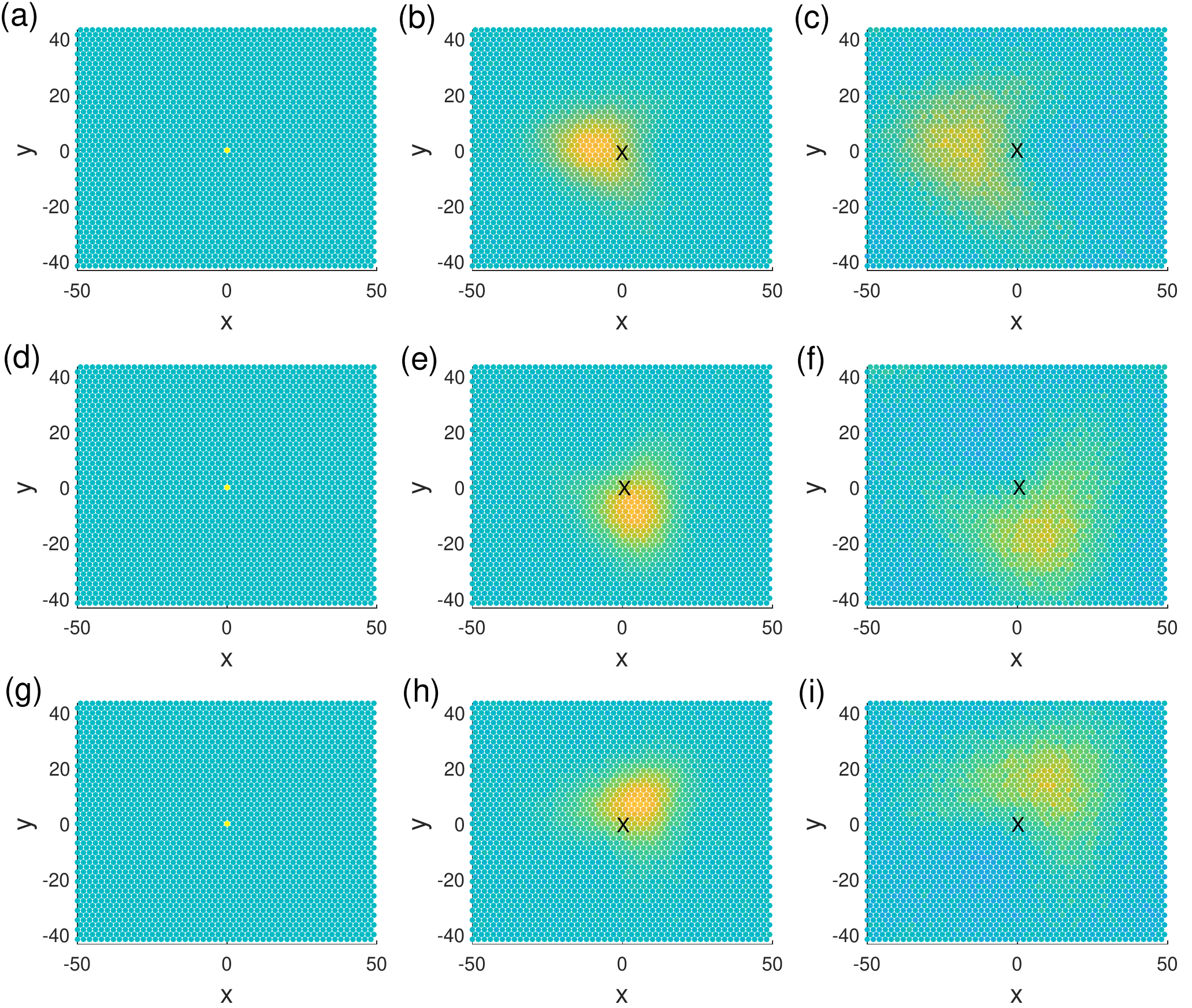}
\caption{The propagating wave implied by the anomalous coefficient in hydrodynamics is captured in numerical simulations of ${C_{ij}(x,y,t)\ e^{(a)}_{i} e^{(a)}_{j}}$.  From top to bottom, each row depicts this correlator for $a=1$, $a=2$, $a=3$ respectively.  The columns of this figure represent different times: from left to right, $t=0$, $t=120$, $t=240$. The origin (0,0) is always marked with an X in the plots for nonzero ${t}$ as a guide to the eye. Simulations here were done with $50\times 50$ unit cells in the lattice.}
\label{propagatingwave}
\end{figure}

We now show the numerical results of large-scale simulations of these Markov chains. The probabilities of acting gate A, B, C are 1/9, 2/9 and 2/3 respectively. We first look for evidence of the sound wave predicted in (\ref{eq:trianglehydroeom}).  The propagating wave can be directly seen from the correlation function ${C_{ij}(x,y,t) = \langle \rho_i(x,y,t) \rho_j(0,0,0) \rangle}$. In Figure \ref{propagatingwave}, we plot $C_{ij} e_{a,i}e_{a,j}$ with the basis vectors $\mathbf{e}_a$ defined in (\ref{eq:ebasis}). For $C_{ij}e_{i,1}e_{j,1}$, there is a propagating wave moving in the negative x-direction; hence, the other two values of $a$ return waves propagating at relative $120^\circ$ angles.

\begin{figure}[t]
\centering
\includegraphics[scale=0.8]{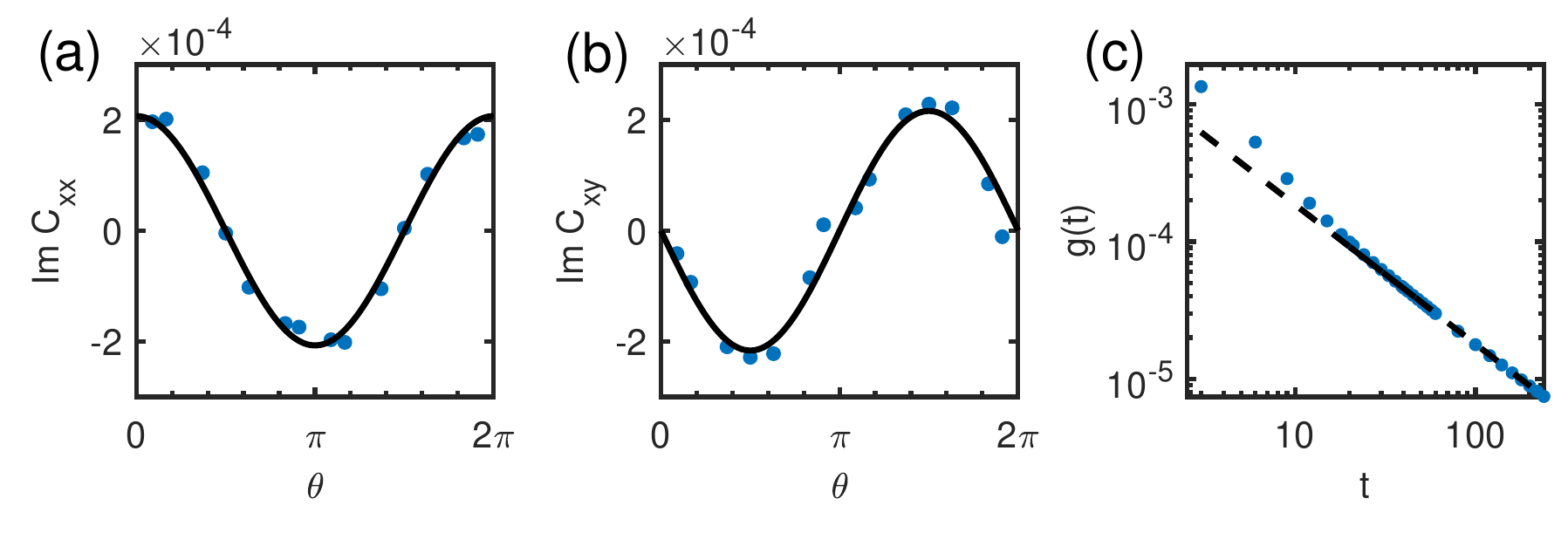}
\caption{(a,b) Confirmation of the triangular hydrodynamic theory by studying the particular structure of the correlation function $C_{ij}(k,t)$, with $k_x=k\cos\theta$ and $k_y=k\sin\theta$.   For given ${|k|}$ and ${t_0}$, the $\theta$-dependence of $C_{ij}$ is given in (\ref{eq:3C}) ${\text{Im}\,C_{xx}(\vec{k},t_0) = \text{Im}\,C_{xx}(\theta,|k|,t_0) \propto \text{cos}(\theta)}$, ${\text{Im}\,C_{xy}(\theta,|k|,t_0) \propto \text{sin}(\theta)}$. The black lines are the theoretical predictions, while the blue dots come from numerical simulations, which were done with $50\times 50$ unit cells in the lattice, at time $t_0=4$, ${|k|=8\pi/L\approx 0.16\pi}$. (c) Algebraic decay in $g(t)$. The dashed line ${\sim t^{-1.01}}$, which is close to the linear response exponent $2/z=1$.  Simulations were done with $100\times 100$ unit cells in the lattice.}
\label{tri}
\end{figure}

 In Figure \ref{tri}, parts (a) and (b), we show that the quantitative structure of the correlation functions in this propagating wave is consistent with our prediction in  (\ref{eq:3C}). To extract the dissipative exponent in the presence of a propagating wave is a bit more subtle. Following \cite{Guo:2022ixk}, we calculate a discretized version of
\begin{eqnarray}\label{gt}
        g(t) \equiv \sum_{ij} \int \mathrm{d}^{2}x \; |C_{ij}(\vec{x},t)|^2 \sim t^{-2/z}.
\end{eqnarray}
Here $z$ is the dynamical critical exponent of the theory. For our system, ${z=2}$ at the hydrodynamic fixed point within linear response theory, and we did not see noteworthy deviations from that prediction.  Indeed, we find $z\approx 2$ in the numerical results shown in Figure \ref{tri} (c). 

\section{Discussion}
In this paper, we have introduced the anomalous hydrodynamics of a theory with vector conserved charge and $\mathrm{D}_6$ symmetry in 2+1 dimensions.  Classical Markov chain simulations have demonstrated that this anomalous hydrodynamics indeed arises in an entirely classical setting, much like the biased random walk.  The effective field theory approach we described allows one to generalize these findings to other point groups, dimensions, and irreducible representations for conserved densities.  

The anomaly of this theory with triangular point group appears to be somewhat unusual.  In conventional field theory, anomalies could not have existed in even spatial dimensions, simply as a consequence of rotational symmetry.  Even at the classical level, the only ``anomalous" terms one could write down involve Levi-Civita tensors, and there is no way to suitably contract indices in 2+1 dimensions.  For the vector conserved charge, this issue has been avoided due to the occurrence of third-rank invariant tensor $\lambda_{ijk}$.  By dimensional analysis, we found an intrinsic length scale arises when analyzing the anomaly, which quantum mechanically could arise from a UV length scale typical of foliated quantum field theories \cite{Slagle_2021}. Curiously, despite being related to a foliated field theory, the quantum mechanical theory analyzed in Appendix \ref{app:A} does not exhibit fractonic behavior along the lines of \cite{Slagle_2021}, instead hosting holomorphic conserved charges as described in Sec. \ref{sec:holomorphicconservation}. The somewhat unexpected connection between this theory and foliated quantum field theory raises the question of what other (non-fractonic) phenomena could be captured within the latter framework.  Alternatively, the length scale could be tied to the size of the system, which would lead to a more subtle manifestation of the UV-IR mixing that arises in theories with exotic symmetry \cite{vectorglobalsym, SeibergShao2d, SeibergShao3d,Gorantla:2020xap,Rudelius:2020kta,subsysteminflow,You:2021tmm,you,Gorantla:2022eem}.  Understanding the length scale $a$ is also interesting, because anomaly coefficients (in this case proportional to $a^{-1}$) are RG-invariant, which is in tension with the naive hydrodynamic scaling dimensions of operators in our classical field theory.   We hope that further analysis on this, and other, anomalous theories in 2+1 dimensions, clarifies the situation in the coming years. 

\section*{Acknowledgments}
 We thank Luca Delacretaz for helpful discussions, and MQ thanks Ho Tat Lam for illuminating discussions on the chiral boson. This work was supported in part by the Alfred P. Sloan Foundation through Grant FG-2020-13795 (AL), the National Science Foundation through CAREER Grant DMR-2145544 (JG, AL), and the NDSEG Fellowship (MQ).

\appendix
\section{Lagrangian free field theory with triangular anomaly and 3+1d anomaly inflow}\label{app:A}
In this appendix, we describe a non-interacting field theory which exhibits the triangular chiral anomaly described in the main text.
\subsection{Warm-up: chiral boson}
We begin, as before, with a brief review of the chiral boson with anomalous U(1), following appendix A of \cite{subsysteminflow}.  Consider a $1+1d$ system described by the real-time action 
\begin{equation} \label{eq:chiralboson}
    S = \frac{N}{4\pi} \int \mathrm{d}x \mathrm{d}t \; \left[ \partial_t \phi \partial_x \phi - v (\partial_x \phi)^2 \right]
\end{equation}
Here $\phi$ is a compact scalar, so $\phi \sim \phi + 2\pi$. One can straightforwardly derive the equations of motion as 
\begin{equation}
    \partial_t (\partial_x \phi) + \partial_x ( \partial_t \phi - 2v \partial_x \phi) = 0,
\end{equation}
which reduces to 
\begin{equation}
    \partial_t \rho - v \partial_x \rho = 0
\end{equation}
after making the identification $\rho \sim \partial_x \phi$. We see that this action reproduces the physics of the biased random walk, at least within linear response.  The non-trivial Poisson bracket (in the classical limit) is encoded via the mixed first term in the action.

Let us now examine the symmetries of the action, which will allow us to justify identifying $\rho$, a conserved charge, with $\partial_x \phi$. The first symmetry we can consider is the shift symmetry of $\phi \mapsto \phi + \alpha$. The Noether current for this symmetry can be found via the usual procedure of allowing $\alpha(x)$ to be spacetime dependent. The corresponding change of the action is 
\begin{equation}
\begin{aligned}
    \delta S &= \frac{N}{4\pi} \int \mathrm{d}x \mathrm{d}t \; \partial_t \alpha \partial_x \phi + \partial_t \phi \partial_x \alpha - 2v \partial_x \alpha \partial_x \phi \\
            &= \frac{N}{4\pi} \int \mathrm{d}x \mathrm{d}t \; 2 \partial_t \alpha \partial_x \phi - 2v \partial_x \alpha \partial_x \phi \\
            &=  \int \mathrm{d}x \mathrm{d}t \; j^t \partial_t \alpha + j^x \partial_x \alpha
\end{aligned}
\end{equation}
so we can identify the current as 
\begin{equation}
    \rho = \frac{N}{2\pi} \partial_x \phi, \;\; J = -v \frac{N}{2\pi} \partial_x \phi.
\end{equation}
The conservation equation reproduces the equation of motion. We can couple the current to a background gauge field by adding to the action a term $-\int \rho A_t + J A_x$, and include a $ \int (A_t - v A_x) A_x$ term for convenience. The full action is 
\begin{equation} \label{eq:chiralbosongaugefield}
    S[A] = \frac{N}{4\pi} \int \mathrm{d}x \mathrm{d}t \; \left[\partial_t \phi \partial_x \phi - v(\partial_x \phi)^2 - 2\partial_x \phi A_t + 2v \partial_x \phi A_x + (A_t - v A_x) A_x\right]
\end{equation}
This action is not invariant under the gauge transformation $\phi \mapsto \phi + \alpha$, $A \mapsto A + \mathrm{d}\alpha$. The action changes by
\begin{equation}
    \delta S[A; \alpha] = \frac{N}{4\pi} \int \mathrm{d}x \mathrm{d}t \; \alpha (\partial_x A_t - \partial_t A_x).
\end{equation}

This lack of gauge invariance signals an anomaly. The anomaly can be cancelled by a bulk Chern-Simons theory (which describes an integer quantum Hall state). Explicitly, this can be shown as follows. Consider the Chern-Simons action 
\begin{equation} \label{eq:bulkCS}
    S_{\mathrm{bulk}}[A] = \frac{N}{4\pi} \intxyt \; \epsilon_{\alpha \beta \gamma} A_\alpha \partial_\beta A_\gamma
\end{equation}
defined on on the region $y \leq 0$. Under a gauge transformation $A_i \to A_i + \partial_i \alpha$, the action changes by 
\begin{equation}
\begin{aligned}
    \delta S_{\mathrm{bulk}}[A;\alpha] &= \frac{N}{4\pi} \intxyt \; \epsilon_{\alpha \beta \gamma} \partial_\alpha \alpha \partial_\beta A_\gamma \\ 
    &= \frac{N}{4\pi} \intxyt \; \partial_\alpha (\epsilon_{\alpha \beta \gamma} \alpha \partial_\beta A_\gamma) \\
    &= \frac{N}{4\pi} \intxt \; \epsilon_{y\beta \gamma} \alpha \partial_\beta A_\gamma \\
    &= \frac{N}{4\pi} \intxt \; \alpha (\partial_t A_x - \partial_x A_t) = -\delta S[A; \alpha].
\end{aligned}
\end{equation}
So the bulk Chern-Simons theory \eqref{eq:bulkCS} together with the boundary \eqref{eq:chiralbosongaugefield} is gauge invariant. Hence, the bulk Chern-Simons theory cancels the anomaly of the boundary via anomaly inflow.  

\subsection{Triangular model}
We now consider a $2+1$-dimensional system given by real-time action 
\begin{equation} \label{eq:trianglelagrangian}
    S = \frac{N}{4\pi a} \int \mathrm{d}x \mathrm{d}y \mathrm{d}t \; \partial_t \phi_i \lambda_{ijk} \partial_j \phi_k - v (\lambda_{ijk} \partial_j \phi_k)^2
\end{equation}
where $\phi_i$ is a two-component compact boson transforming as a vector under the triangle point group, and $a$ is a length scale which is (as of now) undetermined. There is again a shift symmetry $\phi_i \to \phi_i + \alpha_i$ for which we can compute the Noether current by allowing $\alpha_i(x)$ to be spacetime dependent. The corresponding change of the action is 
\begin{equation}
\begin{aligned}
    \delta S &= \frac{N}{4\pi a} \intxyt \;  \partial_t \alpha_i \partial_t \alpha_i \lambda_{ijk} \partial_j \phi_k + \partial_t \phi_i \lambda_{ijk} \partial_j \alpha_k - 2v \lambda_{ijk} \partial_j \phi_k \lambda_{ilm} \partial_l \alpha_m \\
    &= \frac{N}{4\pi a} \intxyt \;  2 \partial_t \alpha_i \lambda_{ijk} \partial_j \phi_k - 2v \partial_j \alpha_i ( \lambda_{jim} \lambda_{mlk} \partial_l \phi_k) \\
    &= \intxyt \; \rho_i \partial_t \alpha_i + J_{ij} \partial_j \alpha_i
\end{aligned}
\end{equation}
so we can identify the conserved charge and current as 
\begin{equation} \label{eq:trianglechargecurrent}
    \rho_i = \frac{N}{2\pi a} \lambda_{ijk} \partial_j \phi_k, \;\; J_{ij} = -v \frac{N}{2\pi a} \lambda_{ijm} \lambda_{mlk} \partial_l \phi_k.
\end{equation}
The continuity equation \begin{eqnarray}
        \partial_t \rho_i + \partial_j J_{ij} = 0
\end{eqnarray}
is the equation of motion. Note that \begin{eqnarray}
        J_{ij} =  -v \lambda_{ijm} \rho_m,
\end{eqnarray}
so this action reproduces the (ideal hydrodynamic) physics of the triangle fluid described in the main text. 

Let us now clarify the issues surrounding compactness of $\phi_i$, the normalization of the action \eqref{eq:trianglelagrangian}, and quantization of the charges $\rho_i$. For consistency with the triangular point group symmetry, we take space to be the torus $\mathbb{R}^2 / \{\Lambda_1, \Lambda_2\}$ where $\Lambda_1 = (L,0)$ and $\Lambda_2 = (L/2, L\sqrt{3}/2)$. This torus carries a natural action of the triangular point group. Similarly, since $\phi_i$ transforms as a vector, we identify 
\begin{equation} \label{eq:phiperiodicity}
    (\phi_x, \phi_y) \sim (\phi_x + 2\pi, \phi_y) \sim (\phi_x + \pi, \phi_y + \pi \sqrt{3}).
\end{equation}
Having established the target space torus, we can now consider winding configurations of $\phi_i$. There are four integer parameters which characterize the possible windings up to homotopy; representative winding configurations are given by 
\begin{equation}
    \label{eq:phiwindings}
    \begin{pmatrix} \phi_x \\ \phi_y \end{pmatrix} = \frac{2\pi}{L}\begin{pmatrix} n_1 + \frac{n_3}{2} & -\frac{2n_1 - 4 n_2 + n_3 - 2n_4}{2 \sqrt{3}} \\  \frac{n_3 \sqrt{3}}{2} & -\frac{n_3}{2} + n_4 \end{pmatrix} \begin{pmatrix} x \\ y \end{pmatrix} 
\end{equation}
with $n_1, n_2, n_3, n_4 \in \mathbb{Z}$. Applying \eqref{eq:trianglechargecurrent}, the total charges are 
\begin{equation} \label{eq:trianglecharges}
\begin{aligned}
    Q_x &= \int \rho_x = (-n_1 - n_3 + n_4) N \frac{L\sqrt{3}}{2a} \\
    Q_y &= \int \rho_y = \frac{-n_1 +2 n_2 + n_3 + n_4}{\sqrt{3}} N \frac{L \sqrt{3}}{2a} .
\end{aligned}
\end{equation}
We take \begin{equation}
    \frac{L \sqrt{3}}{2a} \in \mathbb{Z},
\end{equation} for reasons which will become clear when we discuss anomaly inflow. Interestingly, for $a$ a microscopic length scale, the minimal winding configurations have charge which scale with system size.  However, the requirement for the theory to be well-posed is not as strong, and would allow for $a\sim L$. We see that the normalization of \eqref{eq:trianglelagrangian} ensures that $\mathbf{Q} \cdot \mathbf{e}_a$ are integer valued. 

As before, we can couple the theory to a background gauge field for the symmetry. We add to the action a term $- \int \rho_i A_{ti} + J_{ij} A_{ij}$ and include a $\int \lambda_{ilm} (A_{ti} - v \lambda_{ijk} A_{jk}) A_{lm}$ term for convenience. The action is 
\begin{equation}
\begin{aligned}
    S[A] = \frac{N}{4\pi a} \intxyt \; &\left[\partial_t \phi_i \lambda_{ijk} \partial_j \phi_k - v (\lambda_{ijk} \partial_j \phi_k)^2 - 2\lambda_{ijk} \partial_j \phi_k A_{ti}\right.  \\ 
    +\; & \left.2v \lambda_{ijm} \lambda_{mlk} \partial_l \phi_k A_{ij} + \lambda_{ilm} (A_{ti} - v \lambda_{ijk} A_{jk}) A_{lm}\right].
\end{aligned}
\end{equation}
Under a gauge transformation $\phi_i \mapsto \phi_i + \alpha_i$, $A_{ti} \mapsto \partial_t \alpha_i$, $A_{ij} \mapsto \partial_j \alpha_i$, the action changes as 
\begin{equation}
    \delta S[A; \alpha] = \frac{N}{4\pi a} \intxyt \; \alpha_i (\lambda_{ijk} \partial_j A_{tk} - \lambda_{ijk} \partial_t A_{jk}). \label{eq:deltaStriangle}
\end{equation}

Like before, (\ref{eq:deltaStriangle}) signals an anomaly. To motivate the bulk theory which cancels this anomaly, let us turn to the Markov chain picture of the anomalous triangle fluid. Recall that the Markov chain consists of three species of charges (one for each type of edge on the triangular lattice) and three types of gates: Gate A, which couples the three species of charges; Gate B, which implements a random walk for each species of charge and leads to diffusion; and Gate C, which introduces a bias to the random walk. The key point is that, ignoring for the moment Gate A, the Markov chain resembles three separate infinite stacks of biased random walks, arranged to form a triangular lattice. Making use of the analogy between the biased random walk and the chiral boson, we can conjecture that a bulk formed from a triangular stacking of quantum Hall states (described by Chern-Simons theory) should cancel the anomaly. 

We now make the previous statements precise. The procedure for defining theories with a ``stack" (foliation) structure was laid out in \cite{Slagle_2021}. Consider first a toy example of a stack of Chern-Simons theories stacked such that the normal vector points along the $y$-direction, while the chiral edge modes propagate along the $x$-direction. Following \cite{Slagle_2021}, the appropriate field theory is 
\begin{equation}
    S \sim \int A \wedge \mathrm{d}A \wedge \frac{\mathrm{d}y}{a}
\end{equation}
where $\mathrm{d}y$ is the coordinate one-form in the $y$-direction and $a$ is a length scale which represents the spacing between the stacks. For now, we leave the normalization undetermined; it will later be fixed by the anomaly matching condition. Now, consider three such stacks of Chern-Simons theories, stacked such that the chiral edge modes in the $xy$-plane boundary propagate along the $\mathbf{e}_a$ directions. We therefore have 
\begin{equation} \label{eq:bulkstack}
\begin{aligned}
    S_{\mathrm{bulk}} &\sim \sum_{a = 1}^3  \int A^a \wedge \mathrm{d}A^a \wedge \mathbf{f}^a_\delta \frac{\mathrm{d}x^\delta}{a} \\
    &\sim \sum_{a = 1}^3  \intxyzt \; \frac{1}{a} \epsilon_{\alpha \beta \gamma \delta} A^a_\alpha \partial_\beta A^a_\gamma \mathbf{f}^a_\delta
\end{aligned}
\end{equation}
where $\mathbf{f}^a_A = \epsilon_{BA} \mathbf{e}^a_B$ are vectors orthogonal to the triangular lattice vectors $\mathbf{e}^a$. The three $A^a$ gauge fields are not independent; we make the identification 
\begin{equation}
    A^a_\alpha = \sum_{A = x,y} A_{\alpha A} \mathbf{e}^a_A
\end{equation}
so that \begin{eqnarray}
        \sum_{a=1}^3A^a=0.
\end{eqnarray} In the above and what follows, we use capital letters $A, B, \ldots$ to denote indices which only take values in $x,y$, lowercase letters $i,j,k,\ldots$ to denote all spatial indices, and Greek letters $\alpha, \beta, \ldots$ to denote spacetime indices. Now we can make use of the identity (\ref{eq:lambdaconstruction2}) to rewrite the action as
\begin{equation}
    S_{\mathrm{bulk}} = \frac{N}{4\pi a} \intxyzt\; \epsilon_{\alpha \beta \gamma D}  \lambda_{ABC} A_{\alpha A} \partial_\beta A_{\gamma B} \epsilon_{CD}.
\end{equation}
where we have fixed the normalization of $S_{\mathrm{bulk}}$ to ensure that the bulk cancels the anomaly of the boundary. To see that this is the case, consider the bulk action defined on a region $z \leq 0$. Upon a gauge transformation $A_{\alpha A} \to A_{\alpha A} + \partial_\alpha \alpha_A$, the action changes by 
\begin{equation}
    \begin{aligned}
        \delta S_{\mathrm{bulk}} &= \frac{N}{4\pi a}  \intxyzt\; \epsilon_{\alpha \beta \gamma D} \epsilon_{CD} \lambda_{ABC} \partial_\alpha \alpha_A \partial_\beta A_{\gamma B} \\ 
        &= \frac{N}{4\pi a}  \intxyzt\; \partial_\alpha (\epsilon_{\alpha \beta \gamma D} \epsilon_{CD} \lambda_{ABC} \alpha_A \partial_\beta A_{\gamma B}) \\
        &= \frac{N}{4\pi a}  \intxyt\; \epsilon_{z\beta\gamma D} \epsilon_{CD} \lambda_{ABC} \alpha_A \partial_\beta A_{\gamma B}.
    \end{aligned}
\end{equation}
In the $\epsilon_{z\beta\gamma D}$, only $\beta$ or $\gamma$ can be $t$; expanding out these possibilities gives 
\begin{equation}
\begin{aligned}
    \delta S_{\mathrm{bulk}} &=  \frac{N}{4\pi a}  \intxyt\; \epsilon_{ED} \epsilon_{CD} \lambda_{ABC} \alpha_A \partial_t A_{EB} - \epsilon_{ED} \epsilon_{CD} \lambda_{ABC} \alpha_A \partial_E A_{tB} \\ 
    &= \frac{N}{4\pi a}  \intxyt\; \alpha_A \lambda_{ABC} (\partial_t A_{CB} - \partial_C A_{tB}) 
\end{aligned}
\end{equation}
which is the same anomaly that occurred in the triangle fluid. 

Finally, let us return to the issue of charge quantization. In arguing for quantization of charge in \eqref{eq:trianglecharges}, we claimed that $\frac{L \sqrt{3}}{2a}$ is integer valued. Here we see that an interpretation of this quantity is simply the number of layers of Chern-Simons theories which comprise the stack in \eqref{eq:bulkstack}, which must be an integer. Interestingly, this would imply that the minimal winding configurations \eqref{eq:phiwindings} carry charges which scale with the system size. Such configurations have energy scaling as $\frac{1}{a}$, similar to the field theories discussed in \cite{SeibergShao2d, SeibergShao3d}. However, in \eqref{eq:trianglelagrangian} $a$ simply plays the role of a length scale required by dimensional analysis rather than a lattice regularizer, and so no $a \to 0$ limit is needed. Indeed, there seems to be no formal obstruction to taking $a \sim L$ so long as $\frac{L \sqrt{3}}{2a}$ is integer valued. This would lead to a rather unusual kind of UV-IR mixing, where the UV action contains an anomalously small $1/L$ prefactor. 

\bibliography{thebib}

\begin{thebibliography}{45}%
\makeatletter
\providecommand \@ifxundefined [1]{%
 \@ifx{#1\undefined}
}%
\providecommand \@ifnum [1]{%
 \ifnum #1\expandafter \@firstoftwo
 \else \expandafter \@secondoftwo
 \fi
}%
\providecommand \@ifx [1]{%
 \ifx #1\expandafter \@firstoftwo
 \else \expandafter \@secondoftwo
 \fi
}%
\providecommand \natexlab [1]{#1}%
\providecommand \enquote  [1]{``#1''}%
\providecommand \bibnamefont  [1]{#1}%
\providecommand \bibfnamefont [1]{#1}%
\providecommand \citenamefont [1]{#1}%
\providecommand \href@noop [0]{\@secondoftwo}%
\providecommand \href [0]{\begingroup \@sanitize@url \@href}%
\providecommand \@href[1]{\@@startlink{#1}\@@href}%
\providecommand \@@href[1]{\endgroup#1\@@endlink}%
\providecommand \@sanitize@url [0]{\catcode `\\12\catcode `\$12\catcode
  `\&12\catcode `\#12\catcode `\^12\catcode `\_12\catcode `\%12\relax}%
\providecommand \@@startlink[1]{}%
\providecommand \@@endlink[0]{}%
\providecommand \url  [0]{\begingroup\@sanitize@url \@url }%
\providecommand \@url [1]{\endgroup\@href {#1}{\urlprefix }}%
\providecommand \urlprefix  [0]{URL }%
\providecommand \Eprint [0]{\href }%
\providecommand \doibase [0]{http://dx.doi.org/}%
\providecommand \selectlanguage [0]{\@gobble}%
\providecommand \bibinfo  [0]{\@secondoftwo}%
\providecommand \bibfield  [0]{\@secondoftwo}%
\providecommand \translation [1]{[#1]}%
\providecommand \BibitemOpen [0]{}%
\providecommand \bibitemStop [0]{}%
\providecommand \bibitemNoStop [0]{.\EOS\space}%
\providecommand \EOS [0]{\spacefactor3000\relax}%
\providecommand \BibitemShut  [1]{\csname bibitem#1\endcsname}%
\let\auto@bib@innerbib\@empty
\bibitem [{\citenamefont {Crossley}\ \emph {et~al.}(2017)\citenamefont
  {Crossley}, \citenamefont {Glorioso},\ and\ \citenamefont
  {Liu}}]{crossley_effective_2017}%
  \BibitemOpen
  \bibfield  {author} {\bibinfo {author} {\bibfnamefont {Michael}\ \bibnamefont
  {Crossley}}, \bibinfo {author} {\bibfnamefont {Paolo}\ \bibnamefont
  {Glorioso}}, \ and\ \bibinfo {author} {\bibfnamefont {Hong}\ \bibnamefont
  {Liu}},\ }\bibfield  {title} {\enquote {\bibinfo {title} {Effective field
  theory of dissipative fluids},}\ }\href {\doibase 10.1007/JHEP09(2017)095}
  {\bibfield  {journal} {\bibinfo  {journal} {Journal of High Energy Physics}\
  }\textbf {\bibinfo {volume} {09}},\ \bibinfo {pages} {095} (\bibinfo {year}
  {2017})}\BibitemShut {NoStop}%
\bibitem [{\citenamefont {Haehl}\ \emph {et~al.}(2016)\citenamefont {Haehl},
  \citenamefont {Loganayagam},\ and\ \citenamefont
  {Rangamani}}]{haehl_fluid_2016}%
  \BibitemOpen
  \bibfield  {author} {\bibinfo {author} {\bibfnamefont {Felix~M.}\
  \bibnamefont {Haehl}}, \bibinfo {author} {\bibfnamefont {R.}~\bibnamefont
  {Loganayagam}}, \ and\ \bibinfo {author} {\bibfnamefont {Mukund}\
  \bibnamefont {Rangamani}},\ }\bibfield  {title} {\enquote {\bibinfo {title}
  {The fluid manifesto: emergent symmetries, hydrodynamics, and black holes},}\
  }\href {\doibase 10.1007/JHEP01(2016)184} {\bibfield  {journal} {\bibinfo
  {journal} {Journal of High Energy Physics}\ }\textbf {\bibinfo {volume}
  {01}},\ \bibinfo {pages} {184} (\bibinfo {year} {2016})}\BibitemShut
  {NoStop}%
\bibitem [{\citenamefont {Jensen}\ \emph {et~al.}(2018)\citenamefont {Jensen},
  \citenamefont {Pinzani-Fokeeva},\ and\ \citenamefont {Yarom}}]{Jensen_2018}%
  \BibitemOpen
  \bibfield  {author} {\bibinfo {author} {\bibfnamefont {Kristan}\ \bibnamefont
  {Jensen}}, \bibinfo {author} {\bibfnamefont {Natalia}\ \bibnamefont
  {Pinzani-Fokeeva}}, \ and\ \bibinfo {author} {\bibfnamefont {Amos}\
  \bibnamefont {Yarom}},\ }\bibfield  {title} {\enquote {\bibinfo {title}
  {Dissipative hydrodynamics in superspace},}\ }\href {\doibase
  10.1007/jhep09(2018)127} {\bibfield  {journal} {\bibinfo  {journal} {Journal
  of High Energy Physics}\ }\textbf {\bibinfo {volume} {09}},\ \bibinfo {pages}
  {127} (\bibinfo {year} {2018})}\BibitemShut {NoStop}%
\bibitem [{\citenamefont {Son}\ and\ \citenamefont
  {Sur\'{o}wka}(2009)}]{Son_2009}%
  \BibitemOpen
  \bibfield  {author} {\bibinfo {author} {\bibfnamefont {Dam~T.}\ \bibnamefont
  {Son}}\ and\ \bibinfo {author} {\bibfnamefont {Piotr}\ \bibnamefont
  {Sur\'{o}wka}},\ }\bibfield  {title} {\enquote {\bibinfo {title}
  {Hydrodynamics with triangle anomalies},}\ }\href {\doibase
  10.1103/physrevlett.103.191601} {\bibfield  {journal} {\bibinfo  {journal}
  {Physical Review Letters}\ }\textbf {\bibinfo {volume} {103}},\ \bibinfo
  {pages} {191601} (\bibinfo {year} {2009})}\BibitemShut {NoStop}%
\bibitem [{\citenamefont {Glorioso}\ \emph {et~al.}(2019)\citenamefont
  {Glorioso}, \citenamefont {Liu},\ and\ \citenamefont
  {Rajagopal}}]{Glorioso_2019}%
  \BibitemOpen
  \bibfield  {author} {\bibinfo {author} {\bibfnamefont {Paolo}\ \bibnamefont
  {Glorioso}}, \bibinfo {author} {\bibfnamefont {Hong}\ \bibnamefont {Liu}}, \
  and\ \bibinfo {author} {\bibfnamefont {Srivatsan}\ \bibnamefont
  {Rajagopal}},\ }\bibfield  {title} {\enquote {\bibinfo {title} {Global
  anomalies, discrete symmetries and hydrodynamic effective actions},}\ }\href
  {\doibase 10.1007/jhep01(2019)043} {\bibfield  {journal} {\bibinfo  {journal}
  {Journal of High Energy Physics}\ }\textbf {\bibinfo {volume} {01}},\
  \bibinfo {pages} {043} (\bibinfo {year} {2019})}\BibitemShut {NoStop}%
\bibitem [{\citenamefont {Delacr\'etaz}\ and\ \citenamefont
  {Glorioso}(2020)}]{luca}%
  \BibitemOpen
  \bibfield  {author} {\bibinfo {author} {\bibfnamefont {Luca~V.}\ \bibnamefont
  {Delacr\'etaz}}\ and\ \bibinfo {author} {\bibfnamefont {Paolo}\ \bibnamefont
  {Glorioso}},\ }\bibfield  {title} {\enquote {\bibinfo {title} {Breakdown of
  diffusion on chiral edges},}\ }\href {\doibase
  10.1103/PhysRevLett.124.236802} {\bibfield  {journal} {\bibinfo  {journal}
  {Phys. Rev. Lett.}\ }\textbf {\bibinfo {volume} {124}},\ \bibinfo {pages}
  {236802} (\bibinfo {year} {2020})}\BibitemShut {NoStop}%
\bibitem [{\citenamefont {Cook}\ and\ \citenamefont {Lucas}(2019)}]{caleb}%
  \BibitemOpen
  \bibfield  {author} {\bibinfo {author} {\bibfnamefont {Caleb~Q.}\
  \bibnamefont {Cook}}\ and\ \bibinfo {author} {\bibfnamefont {Andrew}\
  \bibnamefont {Lucas}},\ }\bibfield  {title} {\enquote {\bibinfo {title}
  {Electron hydrodynamics with a polygonal fermi surface},}\ }\href {\doibase
  10.1103/PhysRevB.99.235148} {\bibfield  {journal} {\bibinfo  {journal} {Phys.
  Rev. B}\ }\textbf {\bibinfo {volume} {99}},\ \bibinfo {pages} {235148}
  (\bibinfo {year} {2019})}\BibitemShut {NoStop}%
\bibitem [{\citenamefont {Friedman}\ \emph {et~al.}(2022)\citenamefont
  {Friedman}, \citenamefont {Cook},\ and\ \citenamefont
  {Lucas}}]{Friedman_triangular}%
  \BibitemOpen
  \bibfield  {author} {\bibinfo {author} {\bibfnamefont {Aaron~J.}\
  \bibnamefont {Friedman}}, \bibinfo {author} {\bibfnamefont {Caleb~Q.}\
  \bibnamefont {Cook}}, \ and\ \bibinfo {author} {\bibfnamefont {Andrew}\
  \bibnamefont {Lucas}},\ }\href {\doibase 10.48550/ARXIV.2202.08269} {\enquote
  {\bibinfo {title} {Hydrodynamics with triangular point group},}\ } (\bibinfo
  {year} {2022}),\ \Eprint {http://arxiv.org/abs/2202.08269} {arXiv:2202.08269}
  \BibitemShut {NoStop}%
\bibitem [{\citenamefont {Huang}\ and\ \citenamefont
  {Lucas}(2022)}]{Huang_2022}%
  \BibitemOpen
  \bibfield  {author} {\bibinfo {author} {\bibfnamefont {Xiaoyang}\
  \bibnamefont {Huang}}\ and\ \bibinfo {author} {\bibfnamefont {Andrew}\
  \bibnamefont {Lucas}},\ }\bibfield  {title} {\enquote {\bibinfo {title}
  {Hydrodynamic effective field theories with discrete rotational symmetry},}\
  }\href {\doibase 10.1007/jhep03(2022)082} {\bibfield  {journal} {\bibinfo
  {journal} {Journal of High Energy Physics}\ }\textbf {\bibinfo {volume}
  {03}},\ \bibinfo {pages} {082} (\bibinfo {year} {2022})}\BibitemShut
  {NoStop}%
\bibitem [{\citenamefont {Cook}\ and\ \citenamefont
  {Lucas}(2021)}]{viscometry}%
  \BibitemOpen
  \bibfield  {author} {\bibinfo {author} {\bibfnamefont {Caleb~Q.}\
  \bibnamefont {Cook}}\ and\ \bibinfo {author} {\bibfnamefont {Andrew}\
  \bibnamefont {Lucas}},\ }\bibfield  {title} {\enquote {\bibinfo {title}
  {Viscometry of electron fluids from symmetry},}\ }\href {\doibase
  10.1103/PhysRevLett.127.176603} {\bibfield  {journal} {\bibinfo  {journal}
  {Phys. Rev. Lett.}\ }\textbf {\bibinfo {volume} {127}},\ \bibinfo {pages}
  {176603} (\bibinfo {year} {2021})}\BibitemShut {NoStop}%
\bibitem [{\citenamefont {Varnavides}\ \emph {et~al.}(2020)\citenamefont
  {Varnavides}, \citenamefont {Jermyn}, \citenamefont {Anikeeva}, \citenamefont
  {Felser},\ and\ \citenamefont {Narang}}]{anisoHydro1}%
  \BibitemOpen
  \bibfield  {author} {\bibinfo {author} {\bibfnamefont {Georgios}\
  \bibnamefont {Varnavides}}, \bibinfo {author} {\bibfnamefont {Adam~S.}\
  \bibnamefont {Jermyn}}, \bibinfo {author} {\bibfnamefont {Polina}\
  \bibnamefont {Anikeeva}}, \bibinfo {author} {\bibfnamefont {Claudia}\
  \bibnamefont {Felser}}, \ and\ \bibinfo {author} {\bibfnamefont {Prineha}\
  \bibnamefont {Narang}},\ }\bibfield  {title} {\enquote {\bibinfo {title}
  {Electron hydrodynamics in anisotropic materials},}\ }\href {\doibase
  10.1038/s41467-020-18553-y} {\bibfield  {journal} {\bibinfo  {journal}
  {Nature Communications}\ }\textbf {\bibinfo {volume} {11}},\ \bibinfo {pages}
  {4710} (\bibinfo {year} {2020})}\BibitemShut {NoStop}%
\bibitem [{\citenamefont {Link}\ \emph {et~al.}(2018)\citenamefont {Link},
  \citenamefont {Narozhny}, \citenamefont {Kiselev},\ and\ \citenamefont
  {Schmalian}}]{anisoHydro2}%
  \BibitemOpen
  \bibfield  {author} {\bibinfo {author} {\bibfnamefont {Julia~M.}\
  \bibnamefont {Link}}, \bibinfo {author} {\bibfnamefont {Boris~N.}\
  \bibnamefont {Narozhny}}, \bibinfo {author} {\bibfnamefont {Egor~I.}\
  \bibnamefont {Kiselev}}, \ and\ \bibinfo {author} {\bibfnamefont {J\"org}\
  \bibnamefont {Schmalian}},\ }\bibfield  {title} {\enquote {\bibinfo {title}
  {Out-of-bounds hydrodynamics in anisotropic dirac fluids},}\ }\href {\doibase
  10.1103/PhysRevLett.120.196801} {\bibfield  {journal} {\bibinfo  {journal}
  {Phys. Rev. Lett.}\ }\textbf {\bibinfo {volume} {120}},\ \bibinfo {pages}
  {196801} (\bibinfo {year} {2018})}\BibitemShut {NoStop}%
\bibitem [{\citenamefont {Rao}\ and\ \citenamefont
  {Bradlyn}(2020)}]{anisoHydro3}%
  \BibitemOpen
  \bibfield  {author} {\bibinfo {author} {\bibfnamefont {Pranav}\ \bibnamefont
  {Rao}}\ and\ \bibinfo {author} {\bibfnamefont {Barry}\ \bibnamefont
  {Bradlyn}},\ }\bibfield  {title} {\enquote {\bibinfo {title} {Hall viscosity
  in quantum systems with discrete symmetry: Point group and lattice
  anisotropy},}\ }\href {\doibase 10.1103/PhysRevX.10.021005} {\bibfield
  {journal} {\bibinfo  {journal} {Phys. Rev. X}\ }\textbf {\bibinfo {volume}
  {10}},\ \bibinfo {pages} {021005} (\bibinfo {year} {2020})}\BibitemShut
  {NoStop}%
\bibitem [{\citenamefont {Rao}\ and\ \citenamefont
  {Bradlyn}(2021)}]{anisoHydro4}%
  \BibitemOpen
  \bibfield  {author} {\bibinfo {author} {\bibfnamefont {Pranav}\ \bibnamefont
  {Rao}}\ and\ \bibinfo {author} {\bibfnamefont {Barry}\ \bibnamefont
  {Bradlyn}},\ }\href {\doibase 10.48550/ARXIV.2112.04545} {\enquote {\bibinfo
  {title} {Resolving hall and dissipative viscosity ambiguities via boundary
  effects},}\ } (\bibinfo {year} {2021}),\ \Eprint
  {http://arxiv.org/abs/2112.04545} {arXiv:2112.04545} \BibitemShut {NoStop}%
\bibitem [{\citenamefont {Gromov}\ \emph {et~al.}(2020)\citenamefont {Gromov},
  \citenamefont {Lucas},\ and\ \citenamefont {Nandkishore}}]{Gromov_2020}%
  \BibitemOpen
  \bibfield  {author} {\bibinfo {author} {\bibfnamefont {Andrey}\ \bibnamefont
  {Gromov}}, \bibinfo {author} {\bibfnamefont {Andrew}\ \bibnamefont {Lucas}},
  \ and\ \bibinfo {author} {\bibfnamefont {Rahul~M.}\ \bibnamefont
  {Nandkishore}},\ }\bibfield  {title} {\enquote {\bibinfo {title} {Fracton
  hydrodynamics},}\ }\href {\doibase 10.1103/physrevresearch.2.033124}
  {\bibfield  {journal} {\bibinfo  {journal} {Physical Review Research}\
  }\textbf {\bibinfo {volume} {2}},\ \bibinfo {pages} {033124} (\bibinfo {year}
  {2020})}\BibitemShut {NoStop}%
\bibitem [{\citenamefont {Glorioso}\ \emph {et~al.}(2022)\citenamefont
  {Glorioso}, \citenamefont {Guo}, \citenamefont {Rodriguez-Nieva},\ and\
  \citenamefont {Lucas}}]{breakdown_fracton_hydro}%
  \BibitemOpen
  \bibfield  {author} {\bibinfo {author} {\bibfnamefont {Paolo}\ \bibnamefont
  {Glorioso}}, \bibinfo {author} {\bibfnamefont {Jinkang}\ \bibnamefont {Guo}},
  \bibinfo {author} {\bibfnamefont {Joaquin~F.}\ \bibnamefont
  {Rodriguez-Nieva}}, \ and\ \bibinfo {author} {\bibfnamefont {Andrew}\
  \bibnamefont {Lucas}},\ }\bibfield  {title} {\enquote {\bibinfo {title}
  {Breakdown of hydrodynamics below four dimensions in a fracton fluid},}\
  }\href {\doibase 10.1038/s41567-022-01631-x} {\bibfield  {journal} {\bibinfo
  {journal} {Nature Physics}\ }\textbf {\bibinfo {volume} {18}},\ \bibinfo
  {pages} {912--917} (\bibinfo {year} {2022})}\BibitemShut {NoStop}%
\bibitem [{\citenamefont {Morningstar}\ \emph {et~al.}(2020)\citenamefont
  {Morningstar}, \citenamefont {Khemani},\ and\ \citenamefont
  {Huse}}]{morningstar}%
  \BibitemOpen
  \bibfield  {author} {\bibinfo {author} {\bibfnamefont {Alan}\ \bibnamefont
  {Morningstar}}, \bibinfo {author} {\bibfnamefont {Vedika}\ \bibnamefont
  {Khemani}}, \ and\ \bibinfo {author} {\bibfnamefont {David~A.}\ \bibnamefont
  {Huse}},\ }\bibfield  {title} {\enquote {\bibinfo {title} {Kinetically
  constrained freezing transition in a dipole-conserving system},}\ }\href
  {\doibase 10.1103/PhysRevB.101.214205} {\bibfield  {journal} {\bibinfo
  {journal} {Phys. Rev. B}\ }\textbf {\bibinfo {volume} {101}},\ \bibinfo
  {pages} {214205} (\bibinfo {year} {2020})}\BibitemShut {NoStop}%
\bibitem [{\citenamefont {Feldmeier}\ \emph {et~al.}(2020)\citenamefont
  {Feldmeier}, \citenamefont {Sala}, \citenamefont {De~Tomasi}, \citenamefont
  {Pollmann},\ and\ \citenamefont {Knap}}]{knap2020}%
  \BibitemOpen
  \bibfield  {author} {\bibinfo {author} {\bibfnamefont {Johannes}\
  \bibnamefont {Feldmeier}}, \bibinfo {author} {\bibfnamefont {Pablo}\
  \bibnamefont {Sala}}, \bibinfo {author} {\bibfnamefont {Giuseppe}\
  \bibnamefont {De~Tomasi}}, \bibinfo {author} {\bibfnamefont {Frank}\
  \bibnamefont {Pollmann}}, \ and\ \bibinfo {author} {\bibfnamefont {Michael}\
  \bibnamefont {Knap}},\ }\bibfield  {title} {\enquote {\bibinfo {title}
  {Anomalous diffusion in dipole- and higher-moment-conserving systems},}\
  }\href {\doibase 10.1103/PhysRevLett.125.245303} {\bibfield  {journal}
  {\bibinfo  {journal} {Phys. Rev. Lett.}\ }\textbf {\bibinfo {volume} {125}},\
  \bibinfo {pages} {245303} (\bibinfo {year} {2020})}\BibitemShut {NoStop}%
\bibitem [{\citenamefont {Zhang}(2020)}]{zhang2020}%
  \BibitemOpen
  \bibfield  {author} {\bibinfo {author} {\bibfnamefont {Pengfei}\ \bibnamefont
  {Zhang}},\ }\bibfield  {title} {\enquote {\bibinfo {title} {Subdiffusion in
  strongly tilted lattice systems},}\ }\href {\doibase
  10.1103/PhysRevResearch.2.033129} {\bibfield  {journal} {\bibinfo  {journal}
  {Phys. Rev. Research}\ }\textbf {\bibinfo {volume} {2}},\ \bibinfo {pages}
  {033129} (\bibinfo {year} {2020})}\BibitemShut {NoStop}%
\bibitem [{\citenamefont {Iaconis}\ \emph {et~al.}(2019)\citenamefont
  {Iaconis}, \citenamefont {Vijay},\ and\ \citenamefont
  {Nandkishore}}]{IaconisVijayNandkishore}%
  \BibitemOpen
  \bibfield  {author} {\bibinfo {author} {\bibfnamefont {Jason}\ \bibnamefont
  {Iaconis}}, \bibinfo {author} {\bibfnamefont {Sagar}\ \bibnamefont {Vijay}},
  \ and\ \bibinfo {author} {\bibfnamefont {Rahul}\ \bibnamefont
  {Nandkishore}},\ }\bibfield  {title} {\enquote {\bibinfo {title} {Anomalous
  subdiffusion from subsystem symmetries},}\ }\href {\doibase
  10.1103/PhysRevB.100.214301} {\bibfield  {journal} {\bibinfo  {journal}
  {Phys. Rev. B}\ }\textbf {\bibinfo {volume} {100}},\ \bibinfo {pages}
  {214301} (\bibinfo {year} {2019})}\BibitemShut {NoStop}%
\bibitem [{\citenamefont {Iaconis}\ \emph {et~al.}(2021)\citenamefont
  {Iaconis}, \citenamefont {Lucas},\ and\ \citenamefont
  {Nandkishore}}]{iaconis2021}%
  \BibitemOpen
  \bibfield  {author} {\bibinfo {author} {\bibfnamefont {Jason}\ \bibnamefont
  {Iaconis}}, \bibinfo {author} {\bibfnamefont {Andrew}\ \bibnamefont {Lucas}},
  \ and\ \bibinfo {author} {\bibfnamefont {Rahul}\ \bibnamefont
  {Nandkishore}},\ }\bibfield  {title} {\enquote {\bibinfo {title} {Multipole
  conservation laws and subdiffusion in any dimension},}\ }\href {\doibase
  10.1103/PhysRevE.103.022142} {\bibfield  {journal} {\bibinfo  {journal}
  {Phys. Rev. E}\ }\textbf {\bibinfo {volume} {103}},\ \bibinfo {pages}
  {022142} (\bibinfo {year} {2021})}\BibitemShut {NoStop}%
\bibitem [{\citenamefont {Doshi}\ and\ \citenamefont {Gromov}(2021)}]{doshi}%
  \BibitemOpen
  \bibfield  {author} {\bibinfo {author} {\bibfnamefont {D.}~\bibnamefont
  {Doshi}}\ and\ \bibinfo {author} {\bibfnamefont {A.}~\bibnamefont {Gromov}},\
  }\bibfield  {title} {\enquote {\bibinfo {title} {Vortices as fractons},}\
  }\href {https://doi.org/10.1038/s42005-021-00540-4} {\bibfield  {journal}
  {\bibinfo  {journal} {Communications Physics}\ }\textbf {\bibinfo {volume}
  {4}},\ \bibinfo {pages} {44} (\bibinfo {year} {2021})}\BibitemShut {NoStop}%
\bibitem [{\citenamefont {Feldmeier}\ \emph {et~al.}(2021)\citenamefont
  {Feldmeier}, \citenamefont {Pollmann},\ and\ \citenamefont
  {Knap}}]{knap2021}%
  \BibitemOpen
  \bibfield  {author} {\bibinfo {author} {\bibfnamefont {Johannes}\
  \bibnamefont {Feldmeier}}, \bibinfo {author} {\bibfnamefont {Frank}\
  \bibnamefont {Pollmann}}, \ and\ \bibinfo {author} {\bibfnamefont {Michael}\
  \bibnamefont {Knap}},\ }\bibfield  {title} {\enquote {\bibinfo {title}
  {Emergent fracton dynamics in a nonplanar dimer model},}\ }\href {\doibase
  10.1103/PhysRevB.103.094303} {\bibfield  {journal} {\bibinfo  {journal}
  {Phys. Rev. B}\ }\textbf {\bibinfo {volume} {103}},\ \bibinfo {pages}
  {094303} (\bibinfo {year} {2021})}\BibitemShut {NoStop}%
\bibitem [{\citenamefont {Grosvenor}\ \emph {et~al.}(2021)\citenamefont
  {Grosvenor}, \citenamefont {Hoyos}, \citenamefont {Pe\~na Ben\'\i{}tez},\
  and\ \citenamefont {Sur\'owka}}]{Grosvenor:2021rrt}%
  \BibitemOpen
  \bibfield  {author} {\bibinfo {author} {\bibfnamefont {Kevin~T.}\
  \bibnamefont {Grosvenor}}, \bibinfo {author} {\bibfnamefont {Carlos}\
  \bibnamefont {Hoyos}}, \bibinfo {author} {\bibfnamefont {Francisco}\
  \bibnamefont {Pe\~na Ben\'\i{}tez}}, \ and\ \bibinfo {author} {\bibfnamefont
  {Piotr}\ \bibnamefont {Sur\'owka}},\ }\bibfield  {title} {\enquote {\bibinfo
  {title} {{Hydrodynamics of ideal fracton fluids}},}\ }\href {\doibase
  10.1103/PhysRevResearch.3.043186} {\bibfield  {journal} {\bibinfo  {journal}
  {Phys. Rev. Res.}\ }\textbf {\bibinfo {volume} {3}},\ \bibinfo {pages}
  {043186} (\bibinfo {year} {2021})},\ \Eprint
  {http://arxiv.org/abs/2105.01084} {arXiv:2105.01084 [cond-mat.str-el]}
  \BibitemShut {NoStop}%
\bibitem [{\citenamefont {Osborne}\ and\ \citenamefont
  {Lucas}(2022)}]{osborne}%
  \BibitemOpen
  \bibfield  {author} {\bibinfo {author} {\bibfnamefont {Andrew}\ \bibnamefont
  {Osborne}}\ and\ \bibinfo {author} {\bibfnamefont {Andrew}\ \bibnamefont
  {Lucas}},\ }\bibfield  {title} {\enquote {\bibinfo {title} {Infinite families
  of fracton fluids with momentum conservation},}\ }\href {\doibase
  10.1103/PhysRevB.105.024311} {\bibfield  {journal} {\bibinfo  {journal}
  {Phys. Rev. B}\ }\textbf {\bibinfo {volume} {105}},\ \bibinfo {pages}
  {024311} (\bibinfo {year} {2022})}\BibitemShut {NoStop}%
\bibitem [{\citenamefont {Burchards}\ \emph {et~al.}(2022)\citenamefont
  {Burchards}, \citenamefont {Feldmeier}, \citenamefont {Schuckert},\ and\
  \citenamefont {Knap}}]{Burchards:2022lqr}%
  \BibitemOpen
  \bibfield  {author} {\bibinfo {author} {\bibfnamefont {A.~G.}\ \bibnamefont
  {Burchards}}, \bibinfo {author} {\bibfnamefont {J.}~\bibnamefont
  {Feldmeier}}, \bibinfo {author} {\bibfnamefont {A.}~\bibnamefont
  {Schuckert}}, \ and\ \bibinfo {author} {\bibfnamefont {M.}~\bibnamefont
  {Knap}},\ }\bibfield  {title} {\enquote {\bibinfo {title} {{Coupled
  Hydrodynamics in Dipole-Conserving Quantum Systems}},}\ }\href@noop {} {\
  (\bibinfo {year} {2022})},\ \Eprint {http://arxiv.org/abs/2201.08852}
  {arXiv:2201.08852 [cond-mat.quant-gas]} \BibitemShut {NoStop}%
\bibitem [{\citenamefont {Hart}\ \emph {et~al.}(2022)\citenamefont {Hart},
  \citenamefont {Lucas},\ and\ \citenamefont
  {Nandkishore}}]{quasiconservation}%
  \BibitemOpen
  \bibfield  {author} {\bibinfo {author} {\bibfnamefont {Oliver}\ \bibnamefont
  {Hart}}, \bibinfo {author} {\bibfnamefont {Andrew}\ \bibnamefont {Lucas}}, \
  and\ \bibinfo {author} {\bibfnamefont {Rahul}\ \bibnamefont {Nandkishore}},\
  }\bibfield  {title} {\enquote {\bibinfo {title} {Hidden quasiconservation
  laws in fracton hydrodynamics},}\ }\href {\doibase
  10.1103/physreve.105.044103} {\bibfield  {journal} {\bibinfo  {journal}
  {Physical Review E}\ }\textbf {\bibinfo {volume} {105}},\ \bibinfo {pages}
  {044103} (\bibinfo {year} {2022})}\BibitemShut {NoStop}%
\bibitem [{\citenamefont {Sala}\ \emph {et~al.}(2021)\citenamefont {Sala},
  \citenamefont {Lehmann}, \citenamefont {Rakovszky},\ and\ \citenamefont
  {Pollmann}}]{sala2021dynamics}%
  \BibitemOpen
  \bibfield  {author} {\bibinfo {author} {\bibfnamefont {Pablo}\ \bibnamefont
  {Sala}}, \bibinfo {author} {\bibfnamefont {Julius}\ \bibnamefont {Lehmann}},
  \bibinfo {author} {\bibfnamefont {Tibor}\ \bibnamefont {Rakovszky}}, \ and\
  \bibinfo {author} {\bibfnamefont {Frank}\ \bibnamefont {Pollmann}},\
  }\href@noop {} {\enquote {\bibinfo {title} {Dynamics in systems with
  modulated symmetries},}\ } (\bibinfo {year} {2021}),\ \Eprint
  {http://arxiv.org/abs/2110.08302} {arXiv:2110.08302 [cond-mat.stat-mech]}
  \BibitemShut {NoStop}%
\bibitem [{\citenamefont {Guo}\ \emph {et~al.}(2022)\citenamefont {Guo},
  \citenamefont {Glorioso},\ and\ \citenamefont {Lucas}}]{Guo:2022ixk}%
  \BibitemOpen
  \bibfield  {author} {\bibinfo {author} {\bibfnamefont {Jinkang}\ \bibnamefont
  {Guo}}, \bibinfo {author} {\bibfnamefont {Paolo}\ \bibnamefont {Glorioso}}, \
  and\ \bibinfo {author} {\bibfnamefont {Andrew}\ \bibnamefont {Lucas}},\
  }\href {\doibase 10.48550/ARXIV.2204.06006} {\enquote {\bibinfo {title}
  {Fracton hydrodynamics without time-reversal symmetry},}\ } (\bibinfo {year}
  {2022}),\ \Eprint {http://arxiv.org/abs/2204.06006} {arXiv:2204.06006}
  \BibitemShut {NoStop}%
\bibitem [{\citenamefont {Seiberg}(2020)}]{vectorglobalsym}%
  \BibitemOpen
  \bibfield  {author} {\bibinfo {author} {\bibfnamefont {Nathan}\ \bibnamefont
  {Seiberg}},\ }\bibfield  {title} {\enquote {\bibinfo {title} {Field theories
  with a vector global symmetry},}\ }\href {\doibase
  10.21468/scipostphys.8.4.050} {\bibfield  {journal} {\bibinfo  {journal}
  {{SciPost} Physics}\ }\textbf {\bibinfo {volume} {8}},\ \bibinfo {pages}
  {050} (\bibinfo {year} {2020})}\BibitemShut {NoStop}%
\bibitem [{\citenamefont {Seiberg}\ and\ \citenamefont
  {Shao}(2021)}]{SeibergShao2d}%
  \BibitemOpen
  \bibfield  {author} {\bibinfo {author} {\bibfnamefont {Nathan}\ \bibnamefont
  {Seiberg}}\ and\ \bibinfo {author} {\bibfnamefont {Shu-Heng}\ \bibnamefont
  {Shao}},\ }\bibfield  {title} {\enquote {\bibinfo {title} {Exotic symmetries,
  duality, and fractons in 2$+$1-dimensional quantum field theory},}\ }\href
  {\doibase 10.21468/scipostphys.10.2.027} {\bibfield  {journal} {\bibinfo
  {journal} {{SciPost} Physics}\ }\textbf {\bibinfo {volume} {10}},\ \bibinfo
  {pages} {027} (\bibinfo {year} {2021})}\BibitemShut {NoStop}%
\bibitem [{\citenamefont {Seiberg}\ and\ \citenamefont
  {Shao}(2020)}]{SeibergShao3d}%
  \BibitemOpen
  \bibfield  {author} {\bibinfo {author} {\bibfnamefont {Nathan}\ \bibnamefont
  {Seiberg}}\ and\ \bibinfo {author} {\bibfnamefont {Shu-Heng}\ \bibnamefont
  {Shao}},\ }\bibfield  {title} {\enquote {\bibinfo {title} {Exotic
  {\textdollar}u(1){\textdollar} symmetries, duality, and fractons in
  3$+$1-dimensional quantum field theory},}\ }\href {\doibase
  10.21468/scipostphys.9.4.046} {\bibfield  {journal} {\bibinfo  {journal}
  {{SciPost} Physics}\ }\textbf {\bibinfo {volume} {9}},\ \bibinfo {pages}
  {046} (\bibinfo {year} {2020})}\BibitemShut {NoStop}%
\bibitem [{\citenamefont {Gorantla}\ \emph {et~al.}(2020)\citenamefont
  {Gorantla}, \citenamefont {Lam}, \citenamefont {Seiberg},\ and\ \citenamefont
  {Shao}}]{Gorantla:2020xap}%
  \BibitemOpen
  \bibfield  {author} {\bibinfo {author} {\bibfnamefont {Pranay}\ \bibnamefont
  {Gorantla}}, \bibinfo {author} {\bibfnamefont {Ho~Tat}\ \bibnamefont {Lam}},
  \bibinfo {author} {\bibfnamefont {Nathan}\ \bibnamefont {Seiberg}}, \ and\
  \bibinfo {author} {\bibfnamefont {Shu-Heng}\ \bibnamefont {Shao}},\
  }\bibfield  {title} {\enquote {\bibinfo {title} {{More Exotic Field Theories
  in 3+1 Dimensions}},}\ }\href {\doibase 10.21468/SciPostPhys.9.5.073}
  {\bibfield  {journal} {\bibinfo  {journal} {SciPost Phys.}\ }\textbf
  {\bibinfo {volume} {9}},\ \bibinfo {pages} {073} (\bibinfo {year} {2020})},\
  \Eprint {http://arxiv.org/abs/2007.04904} {arXiv:2007.04904
  [cond-mat.str-el]} \BibitemShut {NoStop}%
\bibitem [{\citenamefont {Rudelius}\ \emph {et~al.}(2021)\citenamefont
  {Rudelius}, \citenamefont {Seiberg},\ and\ \citenamefont
  {Shao}}]{Rudelius:2020kta}%
  \BibitemOpen
  \bibfield  {author} {\bibinfo {author} {\bibfnamefont {Tom}\ \bibnamefont
  {Rudelius}}, \bibinfo {author} {\bibfnamefont {Nathan}\ \bibnamefont
  {Seiberg}}, \ and\ \bibinfo {author} {\bibfnamefont {Shu-Heng}\ \bibnamefont
  {Shao}},\ }\bibfield  {title} {\enquote {\bibinfo {title} {{Fractons with
  Twisted Boundary Conditions and Their Symmetries}},}\ }\href {\doibase
  10.1103/PhysRevB.103.195113} {\bibfield  {journal} {\bibinfo  {journal}
  {Phys. Rev. B}\ }\textbf {\bibinfo {volume} {103}},\ \bibinfo {pages}
  {195113} (\bibinfo {year} {2021})},\ \Eprint
  {http://arxiv.org/abs/2012.11592} {arXiv:2012.11592 [cond-mat.str-el]}
  \BibitemShut {NoStop}%
\bibitem [{\citenamefont {Burnell}\ \emph {et~al.}(2022)\citenamefont
  {Burnell}, \citenamefont {Devakul}, \citenamefont {Gorantla}, \citenamefont
  {Lam},\ and\ \citenamefont {Shao}}]{subsysteminflow}%
  \BibitemOpen
  \bibfield  {author} {\bibinfo {author} {\bibfnamefont {Fiona~J.}\
  \bibnamefont {Burnell}}, \bibinfo {author} {\bibfnamefont {Trithep}\
  \bibnamefont {Devakul}}, \bibinfo {author} {\bibfnamefont {Pranay}\
  \bibnamefont {Gorantla}}, \bibinfo {author} {\bibfnamefont {Ho~Tat}\
  \bibnamefont {Lam}}, \ and\ \bibinfo {author} {\bibfnamefont {Shu-Heng}\
  \bibnamefont {Shao}},\ }\bibfield  {title} {\enquote {\bibinfo {title}
  {Anomaly inflow for subsystem symmetries},}\ }\href {\doibase
  10.1103/physrevb.106.085113} {\bibfield  {journal} {\bibinfo  {journal}
  {Physical Review B}\ }\textbf {\bibinfo {volume} {106}},\ \bibinfo {pages}
  {085113} (\bibinfo {year} {2022})}\BibitemShut {NoStop}%
\bibitem [{\citenamefont {You}\ \emph {et~al.}(2021)\citenamefont {You},
  \citenamefont {Bibo}, \citenamefont {Hughes},\ and\ \citenamefont
  {Pollmann}}]{You:2021tmm}%
  \BibitemOpen
  \bibfield  {author} {\bibinfo {author} {\bibfnamefont {Yizhi}\ \bibnamefont
  {You}}, \bibinfo {author} {\bibfnamefont {Julian}\ \bibnamefont {Bibo}},
  \bibinfo {author} {\bibfnamefont {Taylor~L.}\ \bibnamefont {Hughes}}, \ and\
  \bibinfo {author} {\bibfnamefont {Frank}\ \bibnamefont {Pollmann}},\
  }\bibfield  {title} {\enquote {\bibinfo {title} {{Fractonic critical point
  proximate to a higher-order topological insulator: How does UV blend with
  IR?}}}\ }\href@noop {} {\  (\bibinfo {year} {2021})},\ \Eprint
  {http://arxiv.org/abs/2101.01724} {arXiv:2101.01724 [cond-mat.str-el]}
  \BibitemShut {NoStop}%
\bibitem [{\citenamefont {You}\ and\ \citenamefont {Moessner}(2021)}]{you}%
  \BibitemOpen
  \bibfield  {author} {\bibinfo {author} {\bibfnamefont {Yizhi}\ \bibnamefont
  {You}}\ and\ \bibinfo {author} {\bibfnamefont {Roderich}\ \bibnamefont
  {Moessner}},\ }\href {https://arxiv.org/abs/2106.07664} {\enquote {\bibinfo
  {title} {Fractonic plaquette-dimer liquid beyond renormalization},}\ }
  (\bibinfo {year} {2021}),\ \Eprint {http://arxiv.org/abs/2106.07664}
  {arXiv:2106.07664} \BibitemShut {NoStop}%
\bibitem [{\citenamefont {Gorantla}\ \emph {et~al.}(2022)\citenamefont
  {Gorantla}, \citenamefont {Lam}, \citenamefont {Seiberg},\ and\ \citenamefont
  {Shao}}]{Gorantla:2022eem}%
  \BibitemOpen
  \bibfield  {author} {\bibinfo {author} {\bibfnamefont {Pranay}\ \bibnamefont
  {Gorantla}}, \bibinfo {author} {\bibfnamefont {Ho~Tat}\ \bibnamefont {Lam}},
  \bibinfo {author} {\bibfnamefont {Nathan}\ \bibnamefont {Seiberg}}, \ and\
  \bibinfo {author} {\bibfnamefont {Shu-Heng}\ \bibnamefont {Shao}},\
  }\bibfield  {title} {\enquote {\bibinfo {title} {{Global dipole symmetry,
  compact Lifshitz theory, tensor gauge theory, and fractons}},}\ }\href
  {\doibase 10.1103/PhysRevB.106.045112} {\bibfield  {journal} {\bibinfo
  {journal} {Phys. Rev. B}\ }\textbf {\bibinfo {volume} {106}},\ \bibinfo
  {pages} {045112} (\bibinfo {year} {2022})},\ \Eprint
  {http://arxiv.org/abs/2201.10589} {arXiv:2201.10589 [cond-mat.str-el]}
  \BibitemShut {NoStop}%
\bibitem [{\citenamefont {Qi}\ \emph {et~al.}(2022)\citenamefont {Qi},
  \citenamefont {Hart}, \citenamefont {Friedman}, \citenamefont {Nandkishore},\
  and\ \citenamefont {Lucas}}]{fracton_magneto}%
  \BibitemOpen
  \bibfield  {author} {\bibinfo {author} {\bibfnamefont {Marvin}\ \bibnamefont
  {Qi}}, \bibinfo {author} {\bibfnamefont {Oliver}\ \bibnamefont {Hart}},
  \bibinfo {author} {\bibfnamefont {Aaron~J.}\ \bibnamefont {Friedman}},
  \bibinfo {author} {\bibfnamefont {Rahul}\ \bibnamefont {Nandkishore}}, \ and\
  \bibinfo {author} {\bibfnamefont {Andrew}\ \bibnamefont {Lucas}},\ }\href
  {\doibase 10.48550/ARXIV.2205.05695} {\enquote {\bibinfo {title} {Fracton
  magnetohydrodynamics},}\ } (\bibinfo {year} {2022}),\ \Eprint
  {http://arxiv.org/abs/2205.05695} {arXiv:2205.05695} \BibitemShut {NoStop}%
\bibitem [{\citenamefont {Spohn}(2014)}]{spohn_nonlinear_2014}%
  \BibitemOpen
  \bibfield  {author} {\bibinfo {author} {\bibfnamefont {Herbert}\ \bibnamefont
  {Spohn}},\ }\bibfield  {title} {\enquote {\bibinfo {title} {Nonlinear
  {Fluctuating} {Hydrodynamics} for {Anharmonic} {Chains}},}\ }\href {\doibase
  10.1007/s10955-014-0933-y} {\bibfield  {journal} {\bibinfo  {journal}
  {Journal of Statistical Physics}\ }\textbf {\bibinfo {volume} {154}},\
  \bibinfo {pages} {1191--1227} (\bibinfo {year} {2014})}\BibitemShut {NoStop}%
\bibitem [{\citenamefont {Alvarez-Gaume}\ and\ \citenamefont
  {Witten}(1984)}]{Alvarez-Gaume:1983ihn}%
  \BibitemOpen
  \bibfield  {author} {\bibinfo {author} {\bibfnamefont {Luis}\ \bibnamefont
  {Alvarez-Gaume}}\ and\ \bibinfo {author} {\bibfnamefont {Edward}\
  \bibnamefont {Witten}},\ }\bibfield  {title} {\enquote {\bibinfo {title}
  {{Gravitational Anomalies}},}\ }\href {\doibase 10.1016/0550-3213(84)90066-X}
  {\bibfield  {journal} {\bibinfo  {journal} {Nucl. Phys. B}\ }\textbf
  {\bibinfo {volume} {234}},\ \bibinfo {pages} {269} (\bibinfo {year}
  {1984})}\BibitemShut {NoStop}%
\bibitem [{\citenamefont {Wen}(1995)}]{Wen:1995qn}%
  \BibitemOpen
  \bibfield  {author} {\bibinfo {author} {\bibfnamefont {Xiao-Gang}\
  \bibnamefont {Wen}},\ }\bibfield  {title} {\enquote {\bibinfo {title}
  {{Topological orders and edge excitations in FQH states}},}\ }\href {\doibase
  10.1080/00018739500101566} {\bibfield  {journal} {\bibinfo  {journal} {Adv.
  Phys.}\ }\textbf {\bibinfo {volume} {44}},\ \bibinfo {pages} {405--473}
  (\bibinfo {year} {1995})},\ \Eprint {http://arxiv.org/abs/cond-mat/9506066}
  {arXiv:cond-mat/9506066} \BibitemShut {NoStop}%
\bibitem [{\citenamefont {Sonnenschein}(1988)}]{SONNENSCHEIN1988752}%
  \BibitemOpen
  \bibfield  {author} {\bibinfo {author} {\bibfnamefont {Jacob}\ \bibnamefont
  {Sonnenschein}},\ }\bibfield  {title} {\enquote {\bibinfo {title} {Chiral
  bosons},}\ }\href {\doibase https://doi.org/10.1016/0550-3213(88)90339-2}
  {\bibfield  {journal} {\bibinfo  {journal} {Nuclear Physics B}\ }\textbf
  {\bibinfo {volume} {309}},\ \bibinfo {pages} {752--770} (\bibinfo {year}
  {1988})}\BibitemShut {NoStop}%
\bibitem [{\citenamefont {Levin}\ \emph {et~al.}(2009)\citenamefont {Levin},
  \citenamefont {Wilmer},\ and\ \citenamefont {Peres}}]{levin}%
  \BibitemOpen
  \bibfield  {author} {\bibinfo {author} {\bibfnamefont {D.~A.}\ \bibnamefont
  {Levin}}, \bibinfo {author} {\bibfnamefont {E.}~\bibnamefont {Wilmer}}, \
  and\ \bibinfo {author} {\bibfnamefont {Y.}~\bibnamefont {Peres}},\
  }\href@noop {} {\emph {\bibinfo {title} {Markov Chains and Mixing Times}}}\
  (\bibinfo  {publisher} {AMS},\ \bibinfo {year} {2009})\BibitemShut {NoStop}%
\bibitem [{\citenamefont {Slagle}(2021)}]{Slagle_2021}%
  \BibitemOpen
  \bibfield  {author} {\bibinfo {author} {\bibfnamefont {Kevin}\ \bibnamefont
  {Slagle}},\ }\bibfield  {title} {\enquote {\bibinfo {title} {Foliated quantum
  field theory of fracton order},}\ }\href {\doibase
  10.1103/physrevlett.126.101603} {\bibfield  {journal} {\bibinfo  {journal}
  {Physical Review Letters}\ }\textbf {\bibinfo {volume} {126}},\ \bibinfo
  {pages} {101603} (\bibinfo {year} {2021})}\BibitemShut {NoStop}%
\end{thebibliography}%


%

\end{document}